\documentclass[12pt,reqno]{amsart}

\usepackage{amssymb,amsthm,amsmath,mathrsfs,}
\usepackage{enumerate}
\usepackage{fullpage}
\usepackage{chngpage}

\numberwithin{equation}{section}

\begin{document}

\newtheorem{theorem}{Theorem}[section]
\newtheorem{lemma}[theorem]{Lemma}
\newtheorem{define}[theorem]{Definition}
\newtheorem{remark}[theorem]{Remark}
\newtheorem{corollary}[theorem]{Corollary}
\newtheorem{example}[theorem]{Example}
\newtheorem{assumption}[theorem]{Assumption}
\newtheorem{proposition}[theorem]{Proposition}
\newtheorem{conjecture}[theorem]{Conjecture}

\def\Ref#1{Ref.~\cite{#1}}

\def\i{\mathrm{i}}
\def\Rnum{{\mathbb R}}
\def\const{\text{const.}}

\def\smallbinom#1#2{{\textstyle \binom{#1}{#2}}}
\def\smallsum{\textstyle\sum}

\def\pr{{\rm pr}}
\def\rk{{\rm rank}}
\def\spn{{\rm span}}
\def\dom{{\rm dom}}
\def\ran{{\rm ran}}
\def\coker{{\rm coker}}
\def\hook{\rfloor\,}
\def\smallsum{\textstyle\sum}

\def\X{\mathbf{X}}
\def\Y{\mathbf{Y}}
\def\w{\boldsymbol\omega}
\def\d{\mathrm{d}}
\def\id{\mathrm{id}}
\def\lieder#1{{\mathcal L}_{#1}}
\def\ad{\mathrm{ad}}
\def\t{\mathrm{t}}
\def\div{\text{div}}

\def\scal{{\rm scal.}}
\def\trans{{\rm trans.}}
\def\can{{\rm can.}}
\def\p{{\rm p}}

\def\PQpair#1#2{\langle #1,#2\rangle}

\def\Jsp{\mathrm{J}^{(\infty)}}
\def\Esp#1{{\mathcal{E}_{#1}}}
\def\symmsp{\mathrm{Symm}}
\def\adjsymmsp{\mathrm{AdjSymm}}
\def\multrsp{\mathrm{Multr}}

\def\Rop{{\mathcal R}}
\def\Dop{{\mathcal D}}
\def\Hop{{\mathcal H}}
\def\Jop{{\mathcal J}}

\def\F{{F}}

\def\d{{\mathbf d}}
\def\grad{{\boldsymbol\nabla}}
\def\lapl{{\boldsymbol\Delta}}

\def\upot{\phi}
\def\vpot{\psi}

\def\aux{{\text{aux}}}

\tolerance=50000
\allowdisplaybreaks[3]

\title{Symmetry actions and brackets\\ for adjoint-symmetries.\\ II: Physical examples}

\author{
Stephen C. Anco${}^\dagger$
\\\\\scshape{
D\lowercase{\scshape{epartment}} \lowercase{\scshape{of}} M\lowercase{\scshape{athematics and}} S\lowercase{\scshape{tatistics}}\\
B\lowercase{\scshape{rock}} U\lowercase{\scshape{niversity}}\\
S\lowercase{\scshape{t.}} C\lowercase{\scshape{atharines}}, ON L2S3A1, C\lowercase{\scshape{anada}}
}}

\thanks{${}^\dagger$sanco@brocku.ca}

\begin{abstract}
Symmetries and adjoint-symmetries are two fundamental (coordinate-free) structures of PDE systems. 
Recent work has developed several new algebraic aspects of adjoint-symmetries: 
three fundamental actions of symmetries on adjoint-symmetries; 
a Lie bracket on the set of adjoint-symmetries given by the range of a symmetry action; 
a generalized Noether (pre-symplectic) operator constructed from any non-variational adjoint-symmetry. 
These results are illustrated here by considering 
five examples of physically interesting nonlinear PDE systems ---
nonlinear reaction-diffusion equations, 
Navier-Stokes equations for compressible viscous fluid flow, 
surface-gravity water wave equations,
coupled solitary wave equations, 
and a nonlinear acoustic equation. 
\end{abstract}

\maketitle

\section{Introduction}

Symmetries and conservation laws are fundamental intrinsic (coordinate-free) structures
of a PDE system \cite{Ovs-book,Olv-book,BCA-book}. 
From an algebraic viewpoint, 
the infinitesimal symmetries of a PDE are the solutions of 
the linearization (Frechet derivative) equation
holding on the space of solutions to the PDE. 
Solutions of the adjoint linearization equation, 
holding on the space of solutions to the PDE, 
are called adjoint-symmetries \cite{SarCanCra1987,SarPriCra1990,AncBlu1997a}
and provide a direct link to conservation laws. 
In particular, 
adjoint-symmetries that satisfy a certain variational condition 
represent multipliers which yield conservation laws
\cite{AncBlu1997a,AncBlu2002b,KraVin,Anc-review}. 

A recent study \cite{Anc2022a} has developed several new algebraic aspects of adjoint-symmetries: 
\begin{itemize}
\item
three fundamental actions of symmetries on adjoint-symmetries 
\item
a Lie bracket on the set of adjoint-symmetries given by the range of a symmetry action
\item 
a generalized Noether (pre-symplectic) operator 
constructed from any adjoint-symmetry that is not a multiplier
\end{itemize}
These results have some clear applications for PDE systems. 
Firstly, 
the symmetry actions on adjoint-symmetries 
can be used to produce a new adjoint-symmetry (and hence possibly a conservation law) 
from a known adjoint-symmetry and a known symmetry, 
while the Lie brackets on adjoint-symmetries 
allow a pair of known adjoint-symmetries to generate a new adjoint-symmetry (and hence possibly a conservation law), 
just as a pair of known symmetries can generate a new symmetry
from their commutator. 
Secondly, 
for evolution PDEs, 
adjoint-symmetries can encode a Hamiltonian structure 
through the existence of a symplectic structure constructed from the Noether operator
and a Hamiltonian functional given by a conservation law. 
Thirdly,
a Lie bracket on adjoint-symmetries provides a corresponding bracket structure for
conservation laws, which is a broad generalization of a Poisson bracket 
applicable to non-Hamiltonian systems. 

The present paper will illustrate these main results by considering 
five examples of physically interesting nonlinear PDE systems:\\
(1) coupled nonlinear reaction-diffusion equations
\\
(2) Navier-Stokes equations for compressible viscous fluid flow 
\\
(3) Boussinesq system for surface-gravity water waves 
\\
(4) coupled solitary wave equations
\\
(5) nonlinear acoustic equation 
\\
PDE systems (1), (2), and (4) will be considered in one spatial dimension;
PDE systems (3) and (5) will be considered respectively in two and three spatial dimensions. 

In each example, 
first, the Lie point symmetries and the low-order adjoint-symmetries
will be summarized. 
Second, 
the three actions of the Lie point symmetries on the adjoint-symmetries
will be presented,
and the corresponding adjoint-symmetry commutator brackets
will be obtained. 
Third, 
in examples (1) and (2), 
a correspondence between symmetries and adjoint-symmetries 
will shown to exist in the absence of any local variational structure (Hamiltonian or Lagrangian) for dissipative PDE systems. 
Fourth, 
a Noether (pre-symplectic) operator 
will be shown to arise directly from the symmetry actions in examples (3) to (5),
using an adjoint-symmetry that is not a multiplier. 
In examples (3) and (4), this operator yields a symplectic 2-form 
and a corresponding Hamiltonian structure. 
In example (5), the Noether operator yields a Lagrangian structure. 
These latter examples will illustrate how variational structures are naturally encoded in 
the adjoint-symmetries of non-dissipative PDE systems. 

All of the symmetries and adjoint-symmetries in the examples are obtained by 
solving the determining equations \eqref{symm.deteqn} and \eqref{adjsymm.deteqn}
through a standard method (see \Ref{Olv-book,Anc-review}). 

The rest of the paper is organized as follows. 
A summary of the symmetry actions, bracket structures, and Noether operator 
is provided in section~\ref{sec:summary}. 
Sections~\ref{sec:reactiondiffusion} to~\ref{sec:acoustic}
contain the five examples. 
Concluding remarks are given in section~\ref{sec:remarks}.

\section{Summary of symmetry actions and brackets for adjoint-symmetries}\label{sec:summary}

The mathematical setting will be calculus in jet space \cite{Olv-book},
which is summarized in the appendix of \Ref{Anc2022a}. 
Partial derivatives and total derivatives are denoted using a coordinate notation. 
The Frechet derivative will be denoted by ${}'$. 
Adjoints of total derivatives and linear operators will be denoted by ${}^*$.
Prolongations will be denoted as $\pr$. 
The transpose of a column/row vector and a matrix will be denoted by $\t$. 
Hereafter,
a ``symmetry'' will refer to an infinitesimal symmetry in evolutionary form. 

As explained in \Ref{Anc2022a}, 
some technical conditions related to local solvablity, involutivity, and existence of a solved form for leading derivatives will be assumed on PDE systems,
which are called regular systems \cite{Anc-review}. 
These conditions hold for essentially all PDE systems of interest in physical applications. 
(See \Ref{AncChe} for additional discussion.)

A general treatment of symmetries relevant for the present work can be found in 
\Ref{Olv-book,BCA-book,Anc-review}.

\subsection{Determining equations for symmetries and adjoint-symmetries}

To begin,
the algebraic formulation of determining equations 
for symmetries and adjoint-symmetries 
will be stated for a general (regular) PDE system of order $N$ consisting of $M$ equations
\begin{equation}\label{pde.sys}
G^A(x,u^{(N)}) =0,
\quad
A=1,\ldots,M
\end{equation}
with independent variables $x^i$, $i=1,\ldots,n$, 
and dependent variables $u^\alpha$, $\alpha=1,\ldots,m$. 
$\Esp{G}$ will denote the solution space of the PDE system. 
The coordinate space $(x^i,u^\alpha,u_j^\alpha,\ldots)$ is called the jet space $\Jsp$.

The determining equation for symmetries is given by 
\begin{equation}\label{symm.deteqn}
G'(P)^A|_\Esp{G} =0 
\end{equation}
where $P^\alpha(x,u^{(k)})$ is a set of functions 
representing the components of the symmetry in evolutionary form. 
Geometrically speaking, 
a symmetry is a vector field $\X_P = P^\alpha\partial_{u^\alpha}$ 
whose prolongation is tangent to $\Esp{G}$ in $\Jsp$. 

The adjoint of the symmetry equation \eqref{symm.deteqn} 
is the determining equation for adjoint-symmetries 
\begin{equation}\label{adjsymm.deteqn}
G'{}^*(Q)_\alpha|_\Esp{G} =0 
\end{equation}
where $Q_A(x,u^{(k)})$ is a set of functions 
representing the components of the adjoint-symmetry. 
Its geometrical meaning is that the 1-form $Q_A\d G^A$ in $\Jsp$ 
functionally vanishes on $\Esp{G}$, as discussed in \Ref{AncWan2020a}. 

Off of the solution space $\Esp{G}$,
these determining equations are respectively given by
\begin{equation}\label{symm.deteqn.offsoln}
G'(P)^A = R_P(G)^A
\end{equation}
and
\begin{equation}\label{adjsymm.deteqn.offsoln}
G'{}^*(Q)_\alpha = R_Q(G)_\alpha
\end{equation}
where $R_P=(R_P)^{A\,I}_{B} D_I$ and $R_Q=(R_Q)_{\alpha\,B}^{I} D_I$ 
are some linear differential operator in total derivatives 
whose coefficients $(R_P)^{A\,I}_{B}$ and $(R_Q)_{\alpha\,B}^{I}$ 
are functions that are non-singular on $\Esp{G}$.

Recall that a multiplier is a set of functions $\Lambda_A(x,u^{(k)})$
that are non-singular on $\Esp{G}$ and satisfy
$\Lambda_A G^A = D_i\Psi^i$ off of $\Esp{G}$, 
for some vector function $\Psi^i$ in $\Jsp$. 
This total divergence condition is equivalent to 
\begin{equation}\label{multr.deteqn}
0 = E_{u^\alpha}(\Lambda_A G^A) = \Lambda'{}^*(G)_\alpha + G'{}^*(\Lambda)_\alpha ,
\end{equation}
whereby $\Lambda_A$ is an adjoint-symmetry, 
\begin{equation}
G'{}^*(\Lambda)_\alpha|_\Esp{G} =0 . 
\end{equation}

These equations \eqref{symm.deteqn.offsoln}, \eqref{adjsymm.deteqn.offsoln}, \eqref{multr.deteqn} 
play a key role in formulating the actions of symmetries on adjoint-symmetries.

\subsection{Actions of symmetries on adjoint-symmetries}\label{sec:symmaction}

There are two basic different actions of symmetries on adjoint-symmetries
\cite{Anc2022a,AncWan2021}. 
One action arises geometrically from applying the Lie derivative with respect to a symmetry $\X_P$ 
to the determining equation for adjoint-symmetries, which yields 
\begin{equation}\label{symmaction2.adjsymm}
Q_A\overset{{\X_P}}{\longrightarrow} Q'(P)_A + R_P^*(Q)_A . 
\end{equation}
The other action comes from the adjoint relationship between the determining equation for infinitesimal symmetries and adjoint-symmetries, yielding 
\begin{equation}\label{symmaction1.adjsymm}
Q_A\overset{{\X_P}}{\longrightarrow} R_P^*(Q)_A - R_Q^*(P)_A . 
\end{equation}
Under this action, adjoint-symmetries are mapped into conservation law multipliers.

For adjoint-symmetries that are conservation law multipliers, 
these two actions coincide with the better known action of symmetries on multipliers
\cite{Anc2016,AncKar,Anc2017}. 
Furthermore, 
the difference of the two actions produces a third action 
\begin{equation}\label{symmaction3.adjsymm}
Q_A\overset{{\X_P}}{\longrightarrow} Q'(P)_A + R_Q^*(P)_A  ,
\end{equation}
which has the property that it vanishes on multipliers,
as seen from the multiplier determining equation \eqref{multr.deteqn}.

\subsection{Dual actions and a Noether operator}\label{sec:noetherop}

For any symmetry action $Q_A {\longrightarrow} S_P(Q)_A$, 
there is a dual action 
\begin{equation}\label{S_Q.op}
S_Q(P)_A:= S_P(Q)_A
\end{equation}  
that maps symmetries into adjoint-symmetries. 
The symmetry action \eqref{symmaction3.adjsymm} is distinguished from the other two 
by the property that its dual action can be expressed as a linear operator in total derivatives:
\begin{equation}\label{Jop.symmaction3}
\Jop := S_{Q} = Q' + R_Q^* . 
\end{equation}

This operator \eqref{Jop.symmaction3} maps symmetries into adjoint-symmetries
and thus is a Noether (pre-symplectic) operator. 
Its formal inverse defines a pre-Hamiltonian (inverse Noether) operator
which maps adjoint-symmetries into symmetries.

\subsection{Lie Bracket for adjoint-symmetries}\label{sec:adjsymmbracket}

A dual symmetry action $S_Q$ can be used to construct an associated Lie bracket 
on the subspace of adjoint-symmetries given by the range of the action. 
This yields a homomorphism from the Lie algebra of symmetries into a Lie algebra of adjoint-symmetries. 

Let $\symmsp_G$ and $\adjsymmsp_G$ denote 
the linear spaces of symmetries and adjoint-symmetries. 
The Lie bracket is given by 
\begin{equation}\label{adjsymm.bracket}
{}^Q[Q_1,Q_2]_A := S_Q([S_Q^{-1}Q_1,S_Q^{-1}Q_2])_A 
\end{equation}
on the linear space $S_Q(\symmsp_G)\subseteq\adjsymmsp_G$. 
Note that, since $S_Q^{-1}$ is well defined only modulo $\ker(S_Q)$,
the condition that $\ker(S_Q)$ is an ideal is necessary and sufficient 
for the bracket to be well defined.  
This condition will select a set of adjoint-symmetries $Q_A$ 
that can be used in constructing the bracket. 
If there is more than one such adjoint-symmetry (up to scaling), 
then a natural choice will be to select a $Q_A$ such that $\ran(S_Q)$ is maximal in $\adjsymmsp_G$. 

An alternative way to have the bracket be well defined arises
when the symmetry Lie algebra contains a scaling symmetry. 
If $\ker(S_Q)$ can be characterized as a subspace by its scaling weight, 
then the symmetry Lie algebra possesses an extra structure of 
a direct-sum decomposition as a linear space
\begin{equation}\label{S_Q:decomp}
\symmsp_G = \ker(S_Q)\oplus\coker(S_Q)
\end{equation}
with $\coker(S_Q)$ being defined by having a distinct scaling weight. 
In this situation, 
$S_Q^{-1}$ can be defined as belonging to the subspace $\coker(S_Q)$, 
and hence the bracket will be well defined. 
More generally, 
a scaling decomposition \eqref{S_Q:decomp} in which 
both $\ker(S_Q)$ and $\coker(S_Q)$ are each a direct sum of 
scaling homogeneous subspaces that have no scaling weights in common
is sufficient.

\subsection{Results for evolution PDEs}\label{sec:evolPDEs}

The preceding general results have a further development 
for evolution PDEs
\begin{equation}\label{evol.pde}
u^\alpha_t = g^\alpha(x,u,\partial_x u,\ldots,\partial_x^N u)
\end{equation}
where $x$ now denotes the spatial independent variables $x^i$, $i=1,\ldots,n$,
while $t$ is the time variable. 
Note that, for such a PDE system, 
\begin{equation}
G^\alpha(t,x,u^{(N)}) = u_t^\alpha - g^\alpha(x,u,\partial_x u,\ldots,\partial_x^N u) 
\end{equation}
with $A=\alpha$ for the indices. 

On the solution space $\Esp{G}$, 
all $t$-derivatives of $u^\alpha$ can be eliminated in any expression
through substituting the equation \eqref{evol.pde} and its spatial derivatives. 
Consequently, 
symmetries consist of a set of functions 
$P^\alpha(t,x,u,\partial_x u,\ldots,\partial_x^k u)$
satisfying 
\begin{equation}\label{evol.symm.deteqn}
\partial_t P^\alpha + P'(g)^\alpha - g'(P)^\alpha
= \partial_t P^\alpha + [g,P]^\alpha =0 
\end{equation}
which is the symmetry determining equation \eqref{symm.deteqn} in simplified form
off of $\Esp{G}$.
Hence, 
\begin{equation}\label{evol.R_P}
R_P = P' . 
\end{equation}
Likewise,
adjoint-symmetries consist of a set of functions 
$Q_\alpha(t,x,u,\partial_x u,\ldots,\partial_x^k u)$
satisfying the adjoint equation 
\begin{equation}\label{evol.adjsymm.deteqn}
-(\partial_t Q_\alpha +Q'(g)_\alpha + g'{}^*(Q)_\alpha)
=0 
\end{equation}
which is a simplified form of the determining equation \eqref{evol.adjsymm.deteqn} 
off of $\Esp{G}$. 
Thus, 
\begin{equation}\label{evol.R_Q}
R_Q = -Q' . 
\end{equation}
An equivalent formulation is given by 
\begin{equation}\label{evol.adjsymm.deteqn.alt}
\partial_t Q_\alpha +\{Q,g\}^*_\alpha=0
\end{equation}
in terms of the anti-commutator $\{A,B\}=A'(B) + B'(A)$,
where $\{A,B\}^*=A'{}^*(B) + B'{}^*(A)$. 

The well-known necessary and sufficient condition 
for an adjoint-symmetry to be a conservation law multiplier is 
that its Frechet derivative is self-adjoint 
\begin{equation}\label{evol.multr}
Q'=Q'{}^* ,
\end{equation}
which follows directly from equations \eqref{evol.R_Q} and \eqref{multr.deteqn}. 
Self-adjointness \eqref{evol.multr} is equivalent to the property that 
$Q_\alpha$ is a variational derivative (gradient) 
\begin{equation}
\Lambda_\alpha = E_{u^\alpha}(\Phi) 
\end{equation}
for some function $\Phi(x,u^{(k)})$, $k\geq 0$,
where $E_{u^\alpha}$ is the Euler operator with respect to $u^\alpha$. 

The relations \eqref{evol.R_P} and \eqref{evol.R_Q} give simplified expressions 
for the symmetry actions \eqref{symmaction2.adjsymm} and \eqref{symmaction1.adjsymm}:
\begin{align}
& Q_\alpha \overset{{\X_P}}{\longrightarrow} Q'(P)_\alpha + P'{}^*(Q)_\alpha, 
\label{evol.symmaction2.adjsymm}  
\\
& Q_\alpha \overset{{\X_P}}{\longrightarrow} Q'{}^*(P)_\alpha + P'{}^*(Q)_\alpha ,  
= E_{u^\alpha}(P^\beta Q_\beta) ,
\label{evol.symmaction1.adjsymm}  
\end{align}
which coincide if $Q_\alpha$ is a conservation law multiplier. 
The symmetry action \eqref{symmaction3.adjsymm} is given by
\begin{equation}\label{evol.symmaction3.adjsymm}
Q_\alpha\overset{{\X_P}}{\longrightarrow} Q'(P)_\alpha -Q'{}^*(P)_\alpha , 
\end{equation}
which is trivial if $Q_\alpha$ is a conservation law multiplier. 

For the sequel, indices will be omitted for simplicity of notation wherever it is convenient.

\subsection{Symplectic 2-form}\label{sec:2form}

The Noether operator defined by the symmetry action \eqref{evol.symmaction3.adjsymm} is simply 
\begin{equation}\label{evol.Jop}
\Jop=Q'-Q'{}^* = -\Jop^* , 
\end{equation}
which is skew. 
It gives rise to a 2-form on the linear space of symmetries: 
\begin{equation}\label{evol.2form}
\w_Q(P_1,P_2) = \int_{\Rnum^n} (P_1^\alpha Q'(P_2)_\alpha - P_2^\alpha Q'(P_1)_\alpha)\, d^nx .
\end{equation}
This 2-form is symplectic, namely $\d\w_Q =0$,
as proven in \Ref{Anc2022a}. 

The formal inverse of the Noether operator \eqref{evol.Jop}
defines a pre-Hamiltonian (inverse Noether) operator $\Jop^{-1}$
which maps adjoint-symmetries into symmetries. 
It also formally yields a Poisson bracket defined by 
\begin{equation}\label{evol.PB}
\{\F_1,\F_2\}_{\Jop^{-1}}:= \int_{\Rnum^n} (\delta \F_1/\delta u)\Jop^{-1}(\delta \F_2/\delta u) \, d^nx
\end{equation}
for functionals $\F=\int_{\Rnum^n} f(x,u^{(k)})\,d^nx$,
where $\delta/\delta u$ denotes the variational derivative, 
namely, $\delta\F/\delta u^\alpha = E_{u^\alpha}(f)$. 
In particular, the Jacobi identity for this bracket holds 
as a consequence of closure of the symplectic 2-form
(see \Ref{Anc2022a}, and also \Ref{Olv-book} for a related general result).

\section{Reaction-diffusion system}\label{sec:reactiondiffusion}

Consider a coupled system of mass-conserving reaction-diffusion equations
with quadratic nonlinearities 
\begin{equation}\label{diffus.sys}
u_t = \kappa_1 u_{xx} + \alpha u^p v , 
\quad
v_t = \kappa_2 v_{xx} - \alpha u^p v, 
\end{equation}  
where 
$\kappa_1>0$, $\kappa_2>0$ are the diffusivity coefficients; 
$\alpha$ is a reaction coefficient; 
$p>0$ is an interaction power. 
This evolution system is a simplified model for
two interacting reactive chemicals or ions that are diffusing in a solute, 
or two proteins in a cell with an activator-inhibitor interaction \cite{IshOtsMoc,MorShi}, 
with densities $u(t,x)$ and $v(t,x)$. 
Here the equilibrium concentrations are $u=v=0$.
Note that more general reactivities $\pm(\alpha_1 u -\beta_1 v)^p(\alpha_2 u -\beta_2 v)$ 
can be expressed in the form \eqref{diffus.sys} 
through a linear transformation on $(u,v)$ 
if the coefficient matrix 
$\begin{pmatrix} \alpha_1 & -\beta_1 \\\alpha_2 & -\beta_2 \end{pmatrix}$ 
has a negative determinant. 

Symmetries (in evolutionary form) $P^u\partial_u+P^v\partial_v$ 
are determined by the equations
\begin{subequations}\label{diffus.sys.symm.deteqns}
\begin{align}
( D_t P^u -\kappa_1 D_x^2 P^u -\alpha (pu^{p-1}v P^u  + u^{p} P^v )|_\Esp{G} =0, 
\\
( D_t P^v -\kappa_2 D_x^2 P^v +\alpha (pu^{p-1}v P^u  + u^{p} P^v )|_\Esp{G} =0, 
\end{align}
\end{subequations}  
where $\Esp{G}$ denotes the solution space of the reaction-diffusion system \eqref{diffus.sys}.
Adjoint-symmetries $(Q^u,Q^v)$ are determined by the adjoint equations
\begin{subequations}\label{diffus.sys.adjsymm.deteqns}
\begin{align}
( {-}D_t Q^u -\kappa_1 D_x^2 Q^u +\alpha pu^{p-1}v (Q^v  - Q^u) )|_\Esp{G} =0, 
\\
( {-}D_t Q^v -\kappa_2 D_x^2 Q^v +\alpha u^{p} (Q^v  - Q^u) )|_\Esp{G} =0 . 
\end{align}
\end{subequations}  
Note that, in the general notation \eqref{pde.sys} for PDE systems, 
here 
\begin{equation}
G=(G^u,G^v)=(u_t - \kappa_1 u_{xx} - \alpha u^p v , v_t - \kappa_2 v_{xx} + \alpha u^p v) . 
\end{equation}

A basis for the linear space of Lie point symmetries, with $P=(P^u,P^v)$, 
consists of 
\begin{equation}
P_1 = (u_t,v_t),
\quad
P_2 = (u_x,v_x),
\quad
P_3 = ( u +p t u_t + \tfrac{1}{2}p x u_x,v+ p t v_t + \tfrac{1}{2}p x v_x),
\end{equation}
which represent generators for 
a time-translation, a space-translation, and a scaling. 
Their algebra is given by the non-zero commutators
\begin{equation}\label{diffus.symm.comm}
[P_1,P_3] = -p P_1 ,
\quad
[P_2,P_3] = - \tfrac{1}{2}p P_2 . 
\end{equation}

A basis of the linear space of adjoint-symmetries,
$Q=(Q^u,Q^v)$, is given by 
\begin{equation}
Q_1 = (1,1),
\quad
Q_2 = (x,x),
\end{equation}
which are also multipliers for conservation laws of 
mass $\mathcal M = \int_\Rnum (u+v)\,dx$
and center of mass $\mathcal X = \int_\Rnum x(u+v)\,dx$. 

Consequently, (cf section~\ref{sec:symmaction})
the third symmetry action \eqref{evol.symmaction3.adjsymm} is trivial,
while the other two symmetry actions \eqref{evol.symmaction2.adjsymm} and \eqref{evol.symmaction1.adjsymm} 
are given by the linear operator
\begin{equation}\label{evol.sys.S_P.1st}
S_P(Q) = (E_u(P Q^\t),E_v(P Q^\t)). 
\end{equation}
This action is summarized in Table~\ref{table:diffus.1stsymmaction}. 
Note that, for evaluating the symmetry actions, 
all $t$-derivatives of $u$ and $v$ are replaced through equations \eqref{diffus.sys}.

\begin{table}[h!]
\caption{Reaction-diffusion system: symmetry action \eqref{evol.sys.S_P.1st} on adjoint-symmetries}
\label{table:diffus.1stsymmaction}
\centering
\begin{tabular}{l||c|c|c}
& $P_1$
& $P_2$
& $P_3$
\\
\hline
\hline  
$Q_1$
& $0$
& $0$
& $(1-\tfrac{1}{2}p)Q_1$
\\
\hline
$Q_2$
& $0$
& $-Q_1$
& $(1-p)Q_2$
\\
\end{tabular}
\end{table}

\subsection{Adjoint-symmetry bracket}
The adjoint-symmetry bracket \eqref{adjsymm.bracket} 
arising from this symmetry action 
is defined via the dual operator 
\begin{equation}\label{evol.sys.1stsymmaction.dual.op}
S_Q(P)=(E_u(P Q^\t),E_v(P Q^\t)) . 
\end{equation}
To obtain the maximal domain, namely the whole linear space $\spn(Q_1,Q_2)$, 
one can choose $Q=Q_2$,
whence
\begin{equation}\label{diffus.1stsymmaction.dual.op}
S_{Q_2}(P)=(E_u(x(P^u+P^v)),E_v(x(P^u+P^v))) . 
\end{equation}
Thereby, one has $\ker(S_{Q_2}) = \spn(P_1)$, 
which is an ideal, 
and 
$\ran(S_{Q_2}^{-1}) =\spn(P_2,P_3)$ modulo $\ker(S_{Q_2})$. 
From Table~\ref{table:diffus.1stsymmaction},
one then obtains
\begin{equation}
S_{Q_2}^{-1}(Q_1) = -P_2,
\quad
S_{Q_2}^{-1}(Q_2) = \tfrac{1}{1-p} P_3 ,
\end{equation}
and thus the adjoint-symmetry  bracket 
can be directly computed by 
\begin{equation}\label{diffus.adjsymm.bracket}
{}^{Q_2}[Q_1,Q_2] = S_{Q_2}([-P_2, \tfrac{1}{1-p}P_3]) =  \tfrac{p}{2(1-p)} S_{Q_2}(P_2)
= \tfrac{p}{2(p-1)} Q_1
\end{equation}
through the symmetry commutator \eqref{diffus.symm.comm}. 

This bracket \eqref{diffus.adjsymm.bracket} is a non-trivial Lie bracket. 
It is isomorphic to the symmetry subalgebra $\mathcal A=\spn(P_2,P_3)$,
which is generated by space translation and scaling. 
This correspondence between symmetries and adjoint-symmetries 
exists in the absence of any local variational structure (Hamiltonian or Lagrangian) 
for the reaction-diffusion equations \eqref{diffus.sys}. 

\begin{table}[h!]
\caption{Reaction-diffusion system: adjoint-symmetry  bracket}
\label{table:diffus.1stadjsymmbracket}
\centering
\begin{tabular}{l||c|c}
& $Q_1$
& $Q_2$
\\
\hline
\hline  
$Q_1$
& $0$
& $\tfrac{p}{2(p-1)} Q_1$
\\
\hline
$Q_2$
& 
& $0$
\\
\end{tabular}
\end{table}

Since the third symmetry action is trivial, 
both the corresponding Noether operator \eqref{evol.Jop} 
and symplectic 2-form \eqref{evol.2form}
are trivial. 
This is expected, as reaction-diffusion systems are inherently dissipative 
(namely, the linearized system is parabolic).

\section{Navier-Stokes equations}\label{sec:NS}

Consider the Navier-Stokes equations for compressible fluid flow \cite{Bat,ChoMar} 
with fluid velocity $u(t,x)$ and the density $\rho(t,x)$ in one spatial dimension
\begin{equation}\label{NS.sys}
u_t +u u_x = (1/\rho) (-p_x + \mu u_{xx}), 
\quad
\rho_t +(u\rho)_x = 0 , 
\end{equation}  
where $\mu\neq 0$ is the viscosity coefficient. 
Here the pressure will be specified by a general polytropic equation of state
\begin{equation}\label{NS.pressure}
p(\rho) = k\rho^q,
\quad
q\neq 0
\end{equation}  
with coefficient $k>0$. 
All of the parameters will be treated as being arbitrary. 
In the general notation \eqref{pde.sys} for PDE systems, 
\begin{equation}
G=(G^u,G^\rho)=(u_t +u u_x + (1/\rho) (p_x - \mu u_{xx}), \rho_t +(u\rho)_x) . 
\end{equation}

The determining equations for symmetries
$P^u\partial_u + P^\rho\partial_\rho$ (in evolutionary form)
are given by 
\begin{subequations}\label{NS.symm.deteqns}
\begin{align}
( D_t P^u + D_x(u P^u) + q D_x((p/\rho^2) P^\rho) +\mu (1/\rho)^2 u_{xx} P^\rho -\mu (1/\rho) D_x^2 P^u )|_\Esp{G} =0, 
\\
( D_t P^\rho + D_x(u P^\rho + \rho P^u) )|_\Esp{G} =0, 
\end{align}
\end{subequations}  
where $\Esp{G}$ denotes the solution space of system \eqref{NS.sys}--\eqref{NS.pressure}.
The adjoint of these equations yields the determining equations for adjoint-symmetries
$(Q^u,Q^\rho)$: 
\begin{subequations}\label{NS.adjsymm.deteqns}
\begin{align}
( {-}D_t Q^u - u D_x Q^u -\rho D_x Q^\rho -\mu D_x^2( (1/\rho) Q^u)  )|_\Esp{G} =0, 
\\
( {-}Q_t P^\rho -u D_x Q^\rho - q(p/\rho^2) D_x Q^u +\mu (1/\rho)^2 u_{xx} Q^u )|_\Esp{G} =0 .
\end{align}
\end{subequations}  

A basis for the linear space of Lie point symmetries, with $P=(P^u,P^\rho)$,
consists of generators for
a time-translation, a space-translation, a Galilean boost, and a scaling:
\begin{equation}
\begin{gathered}
P_1 = (u_t,\rho_t),
\quad
P_2 = (u_x,\rho_x),
\quad
P_3 = (1-t u_x,-t \rho_x),
\\
P_4 = (\tfrac{1-q}{1+q} u -\tfrac{2q}{1+q} t u_t -x u_x,-\tfrac{2}{1+q} \rho -\tfrac{2q}{1+q} t \rho_t -x \rho_x) . 
\end{gathered}
\end{equation}
Their algebra is given by the non-zero commutators
\begin{equation}\label{NS.symm.comm}
[P_1,P_3] = P_2 ,
\quad
[P_1,P_4] = \tfrac{2q}{1+q} P_1 ,
\quad
[P_2,P_4] = P_2,
\quad
[P_3,P_4] = \tfrac{1-q}{1+q} P_3 .
\end{equation}

A basis of the linear space of adjoint-symmetries,
$Q=(Q^u,Q^\rho)$, is given by 
\begin{equation}
Q_1 = (0,1),
\quad
Q_2 = (\rho,u),
\quad
Q_3 = (t\rho,t u -x) . 
\end{equation}
They are multipliers for conservation laws of 
mass $\mathcal M = \int_\Rnum \rho\,dx$, 
momentum $\mathcal P = \int_\Rnum \rho u\,dx$, 
and Galilean momentum $\mathcal G = \int_\Rnum \rho(t u - x)\,dx = t\mathcal P -\mathcal X(t)$
which is related to the motion of the center of mass 
$\mathcal X(t) = \int_\Rnum x\rho\,dx$. 

The third symmetry action \eqref{evol.symmaction3.adjsymm} is trivial
(cf section~\ref{sec:symmaction}), 
while the other two symmetry actions \eqref{evol.symmaction1.adjsymm}, \eqref{evol.symmaction2.adjsymm}
are given by the linear operator \eqref{evol.sys.S_P.1st} with $v=\rho$, 
\begin{equation}\label{NS.S_P}
S_P(Q) = (E_u(P Q^\t),E_\rho(P Q^\t)). 
\end{equation}
This action is summarized in Table~\ref{table:NS.1stsymmaction}. 
For evaluating the symmetry actions, 
all $t$-derivatives of $u$ and $\rho$ are replaced through equations \eqref{NS.sys}. 

\begin{table}[h!]
\caption{Navier-Stokes equations: symmetry action \eqref{NS.S_P} on adjoint-symmetries}
\label{table:NS.1stsymmaction}
\centering
\begin{tabular}{l||c|c|c|c}
& $P_1$
& $P_2$
& $P_3$
& $P_4$
\\
\hline
\hline  
$Q_1$
& $0$
& $0$
& $0$
& $\tfrac{q-1}{q+1}Q_1$
\\
\hline
$Q_2$
& $0$
& $0$
& $Q_1$
& $0$
\\
\hline
$Q_3$
& $-Q_2$
& $Q_1$
& $0$
& $\tfrac{2q}{q+1} Q_3$
\\
\end{tabular}
\end{table}

\subsection{Adjoint-symmetry bracket}
To obtain a maximal domain on which an adjoint-symmetry  bracket \eqref{adjsymm.bracket} 
can be defined via the symmetry action \eqref{NS.S_P}, 
one seeks a maximal range for the dual operator 
\begin{equation}\label{NS.S_Q}
S_Q(P)=(E_u(P Q^\t),E_\rho(P Q^\t)) .
\end{equation}
From Table~\ref{table:NS.1stsymmaction},
it is clear that the maximal range will be the whole linear space of adjoint-symmetries,
which is obtained for the choice $Q=Q_3+c_2 Q_2 +c_1Q_1$,
where one has 
\begin{equation}\label{NS.ker.S_Q}
\ker(S_{Q_3+c_2 Q_2 +c_1 Q_1}) = \spn(P_3-c_2 P_2) .
\end{equation}

The subalgebra \eqref{NS.ker.S_Q} is not an ideal,
since $[P_1,P_3-c_2 P_2] = P_2$,
and consequently the resulting adjoint-symmetry bracket will depend on how 
$\coker(S_{Q_3+c_2Q_2+c_1Q_1})$ is chosen in $\spn(P_1,P_2,P_3,P_4)$. 
However, the scaling symmetry $P_4$ can be utilized 
(cf section~\ref{sec:adjsymmbracket})
to fix a canonical choice of $\coker(S_{Q_3+c_2Q_2+c_1Q_1})$ as follows. 
From the symmetry commutators \eqref{NS.symm.comm},
observe that $\spn(P_3)$ has a different scaling weight compared to 
$\spn(P_1)$ and $\spn(P_2)$. 
Also observe that $\spn(Q_1)$, $\spn(Q_2)$, and $\spn(Q_3)$ have different scaling weights. 
Hence, one can take $c_2=c_1=0$, whereby 
$\ker(S_{Q_3}) = \spn(P_3)$ 
and $\coker(S_{Q_3}) = \spn(P_4)\oplus\spn(P_1)\oplus\spn(P_2)$
provides a well-defined decomposition \eqref{S_Q:decomp}
of the symmetry Lie algebra under scaling. 
This determines $\ran(S_{Q_3}^{-1}) =\spn(P_1,P_2,P_4)$. 

From Table~\ref{table:NS.1stsymmaction},
one now obtains
\begin{equation}\label{NS.inv.S_Q}
S_{Q_3}^{-1}(Q_1) = P_2,
\quad
S_{Q_3}^{-1}(Q_2) = -P_1,
\quad
S_{Q_3}^{-1}(Q_3) = \tfrac{q+1}{2q} P_4 .
\end{equation}
Hence,
the adjoint-symmetry  bracket \eqref{adjsymm.bracket} can be directly computed by 
\begin{subequations}\label{NS.adjsymm.bracket}
\begin{align}
{}^{Q_3}[Q_1,Q_2] & 
= S_{Q_3}([P_2, -P_1]) =  S_{Q_3}(0) = 0, 
\\
{}^{Q_3}[Q_1,Q_3] & 
= S_{Q_3}([P_2, \tfrac{q+1}{2q}P_4]) =  \tfrac{q+1}{2q} S_{Q_3}(P_2) = \tfrac{q+1}{2q} Q_1 , 
\\
{}^{Q_3}[Q_2,Q_3] & 
= S_{Q_3}([-P_1, \tfrac{q+1}{2q}P_4]) =  -S_{Q_3}(P_1) =Q_2 , 
\end{align}
\end{subequations}
through the symmetry commutator \eqref{NS.symm.comm}. 

This bracket \eqref{NS.adjsymm.bracket} is a non-trivial Lie bracket
on the whole linear space $\spn(Q_1,Q_2,Q_3)$ of adjoint-symmetries. 
In particular, from the inverse of the dual operator \eqref{NS.inv.S_Q}, 
one sees that this Lie bracket structure is isomorphic to the symmetry subalgebra 
$\mathcal A=\spn(P_1,P_2,P_4)$, 
which is generated by time translation, space translation, and scaling. 

Remarkably, this correspondence between symmetries and adjoint-symmetries 
exists despite the lack of any local variational structure (Hamiltonian or Lagrangian) 
for the Navier-Stokes equations \eqref{NS.sys}. 

\begin{table}[h!]
\caption{Navier-Stokes equations: adjoint-symmetry  bracket}
\label{table:NS.1stadjsymmbracket}
\centering
\begin{tabular}{l||c|c|c}
& $Q_1$
& $Q_2$
& $Q_3$
\\
\hline
\hline  
$Q_1$
& $0$
& $0$
& $\tfrac{q+1}{2q} Q_1$
\\
\hline
$Q_2$
& 
& $0$
& $Q_2$
\\
\hline
$Q_3$
&
&
& $0$
\\
\end{tabular}
\end{table}

Finally, since the third symmetry action is trivial, 
both the corresponding Noether operator \eqref{evol.Jop} 
and symplectic 2-form \eqref{evol.2form}
are trivial. 
This is expected, as viscous fluid flow is inherently dissipative.

\section{Boussinesq system}\label{sec:boussinesq}

Long wavelength, small amplitude surface water waves
are described by a system of coupled Boussinesq equations 
\cite{BonCheSau,BonColLan}
\begin{equation}
\vec{v}_t + \nabla h + \vec{v}\cdot\nabla\vec{v} -\Delta \vec{v}_t =0,
\quad
h_t +\nabla\cdot\vec{v} +\nabla\cdot(h\vec{v}) - \Delta h_t=0
\end{equation}
where, up to scaling of variables, 
$h(t,x,y)$ is the water elevation and $\vec{v}(t,x,y)$ is the horizontal velocity of the water. 
For irrotational flow, one has $\vec{v}=\nabla u$, 
which gives the coupled equations
\begin{equation}\label{boussinesq.sys}
u_t + h +\tfrac{1}{2} |\nabla u|^2 -\Delta u_t =0,
\quad
h_t +\Delta u +\nabla\cdot(h\nabla u) - \Delta h_t=0
\end{equation}
for $u(t,x,y)$ and $h(t,x,y)$. 
This system is denoted as $G=(G^u,G^h)$
in the general notation \eqref{pde.sys} for PDE systems. 

Symmetries (in evolutionary form) $P^u\partial_u+P^h\partial_h$ 
are determined by the equations
\begin{subequations}\label{boussinesq.sys.symm.deteqns}
\begin{align}
( D_t(1 - D_x^2 -D_y^2) P^u   + u_x D_x P^u + u_y D_y P^u + P^h  )|_\Esp{G} =0, 
\\
\begin{aligned}
( D_t(1 - D_x^2 -D_y^2) P^h   + D_x(u_x P^h + h D_x P^u)+ D_y(u_y P^h + h D_y P^u)& \\
+ (D_x^2+D_y^2) P^u )|_\Esp{G} &=0, 
\end{aligned}
\end{align}
\end{subequations}  
where $\Esp{G}$ denotes the solution space of the system \eqref{boussinesq.sys}.
The linear space of Lie point symmetries, with $P=(P^u,P^h)$, 
is generated by a shift, a time-translation, space-translations, a rotation, and a scaling:
\begin{equation}\label{boussinesq.sys.symms}
\begin{aligned}
&
P_1 = (1,0), 
\quad
P_2 = (u_t,h_t),
\quad
P_3= (u_x,h_x),
\quad
P_4= (u_y,h_y),
\\
& 
P_5 =(x u_y - y u_x,x h_y - y h_x),
\quad
P_6 = (u + t(u_t -2),2h + t h_t + 2) .
\end{aligned}
\end{equation}
Their algebra has the non-zero commutators 
\begin{equation}\label{boussinesq.sys.symm.alg}
\begin{gathered}\hfil
[P_1,P_6] = P_1, 
\quad
[P_2,P_6] = 2P_1 - P_2 ,
\quad
[P_3,P_5] = -P_4,
\quad
[P_4,P_5] = P_3 . 
\end{gathered}  
\end{equation}

Adjoint-symmetries $Q=(Q^u,Q^h)$ are determined by the adjoint of the symmetry equations
\begin{subequations}\label{boussinesq.sys.adjsymm.deteqns}
\begin{align}
\begin{aligned}
( (D_x^2 +D_y^2 -1)D_t Q^u   -D_x(u_x Q^u) -D_y(u_y Q^u) + (D_x^2 +D_y^2) Q^h  &\\
+ D_x(h D_x Q^h) + D_y(h D_y Q^h)  )|_\Esp{G} & =0, 
\end{aligned}
\\
( (D_x^2 +D_y^2 -1)D_t Q^h   -u_x D_x Q^h -u_y D_y Q^h +Q^u )|_\Esp{G} =0 .
\end{align}
\end{subequations}  
The linear space of low-order adjoint-symmetries 
is given by the basis
\begin{equation}\label{boussinesq.sys.adjsymms}
\begin{aligned}
&
Q_1 = (1,0),
\quad
Q_2= (h_t,-u_t), 
\quad
Q_3 =(h_x,-u_x), 
\quad
Q_4 =(h_y,-u_y), 
\\
& 
Q_5 =(x h_y - y h_x,y u_x - x u_y),
\quad
Q_6 = (-2h - t h_t - 2) ,u + t(u_t -2)) .
\end{aligned}
\end{equation}
The first five of these adjoint-symmetries are also multipliers for conservation laws of 
(up to an overall factor) 
mass, $x$- and $y$- momenta, angular momentum, and energy:
\begin{subequations}
\begin{align}
\mathcal M & = \int_{\Rnum^2} h\,dx\,dy,
\\
\mathcal P^x & = \int_{\Rnum^2} ( u_x h_{xx} + u_y h_{xy} + u h_x ) \,dx\,dy, 
\quad
\mathcal P^y = \int_{\Rnum^2} ( u_x h_{xy} + u_y h_{yy} + u h_y ) \,dx\,dy,
\\
\mathcal I & = \int_{\Rnum^2} ( y (u_x h_{xx} + u_y h_{xy} + u h_x) -x(u_x h_{xy} + u_y h_{yy} + u h_y) )\,dx\,dy,
\\
\mathcal E & = \int_{\Rnum^2} \tfrac{1}{2}( h^2 + h(u_x^2 +u_y^2) )\,dx\,dy . 
\end{align}
\end{subequations}
The sixth adjoint-symmetry is not a multiplier. 
Consequently, (cf section~\ref{sec:symmaction})
all three symmetry actions \eqref{symmaction2.adjsymm}, \eqref{symmaction1.adjsymm}, \eqref{symmaction3.adjsymm}
on adjoint-symmetries are different. 

Moreover, the first and second actions differ only on the non-multiplier $Q_6$. 
They are computed by use of: 
\begin{equation}
\begin{aligned}
& R_{P_1} = 0,
\quad
R_{P_2} =  \begin{pmatrix} D_t & 0 \\ 0 & D_t \end{pmatrix},
\quad
R_{P_3} = \begin{pmatrix} D_x & 0 \\ 0 & D_x \end{pmatrix},
\quad
R_{P_4} = \begin{pmatrix} D_y & 0 \\ 0 & D_y \end{pmatrix},
\\
& R_{P_5} = \begin{pmatrix} x D_y - y D_x & 0 \\ 0 & x D_y - y D_x \end{pmatrix},
\quad
R_{P_6} = \begin{pmatrix} tD_t +2 & 0 \\ 0 & tD_t +3 \end{pmatrix} , 
\end{aligned}
\end{equation}
and 
\begin{equation}
\begin{aligned}
& R_{Q_1} = 0,
\quad
R_{Q_2} =  \begin{pmatrix} 0 & D_t \\ -D_t & 0 \end{pmatrix},
\quad
R_{Q_3} = \begin{pmatrix} 0 & D_x \\ -D_x  & 0 \end{pmatrix},
\quad
R_{Q_4} = \begin{pmatrix} 0 & D_y \\ -D_y & 0 \end{pmatrix},
\\
& R_{Q_5} = \begin{pmatrix} 0 & x D_y - y D_x \\ y D_x - x D_y & 0 \end{pmatrix},
\quad
R_{Q_6} = \begin{pmatrix} 0 & tD_t +3 \\ -tD_t -2 & 0 \end{pmatrix} , 
\end{aligned}
\end{equation}
which are readily derived from the expressions \eqref{boussinesq.sys.symms} and \eqref{boussinesq.sys.adjsymms}. 
This leads to the results shown in Tables~\ref{table:boussinesqsys.2ndsymmaction} and~\ref{table:boussinesqsys.1stsymmaction}. 

\begin{table}[h!]
\caption{Boussinesq system: symmetry action \eqref{symmaction2.adjsymm} on adjoint-symmetries}
\label{table:boussinesqsys.2ndsymmaction}
\centering
\begin{tabular}{l||c|c|c|c|c|c}
& $P_1$
& $P_2$
& $P_3$
& $P_4$
& $P_5$
& $P_6$
\\
\hline
\hline  
$Q_1$
& $0$
& $0$
& $0$
& $0$
& $0$
& $2Q_1$
\\
\hline
$Q_2$
& $0$
& $0$
& $0$
& $0$
& $0$
& $4Q_2 -2Q_1$
\\
\hline
$Q_3$
& $0$
& $0$
& $0$
& $0$
& $Q_4$
& $3Q_3$
\\
\hline
$Q_4$
& $0$
& $0$
& $0$
& $0$
& $-Q_3$
& $3Q_4$
\\
\hline
$Q_5$
& $0$
& $0$
& $-Q_4$
& $Q_3$
& $0$
& $3Q_5$
\\
\hline
$Q_6$
& $Q_1$
& $2Q_1-Q_2$
& $0$
& $0$
& $0$
& $3Q_6$
\\
\end{tabular}
\end{table}

\begin{table}[h!]
\caption{Boussinesq system: symmetry action \eqref{symmaction1.adjsymm} on adjoint-symmetries}
\label{table:boussinesqsys.1stsymmaction}
\centering
\begin{tabular}{l||c|c|c|c|c|c}
& $P_1$
& $P_2$
& $P_3$
& $P_4$
& $P_5$
& $P_6$
\\
\hline
\hline  
$Q_1$
& $0$
& $0$
& $0$
& $0$
& $0$
& $2Q_1$
\\
\hline
$Q_2$
& $0$
& $0$
& $0$
& $0$
& $0$
& $4Q_2 -2Q_1$
\\
\hline
$Q_3$
& $0$
& $0$
& $0$
& $0$
& $Q_4$
& $3Q_3$
\\
\hline
$Q_4$
& $0$
& $0$
& $0$
& $0$
& $-Q_3$
& $3Q_4$
\\
\hline
$Q_5$
& $0$
& $0$
& $-Q_4$
& $Q_3$
& $0$
& $3Q_5$
\\
\hline
$Q_6$
& $-2Q_1$
& $2Q_1-4Q_2$
& $-3Q_3$
& $-3Q_4$
& $-3Q_5$
& $0$
\\
\end{tabular}
\end{table}

Using the first action \eqref{symmaction2.adjsymm}, 
one can obtain an adjoint-symmetry bracket on a maximal domain 
which is given by having a maximal range for the dual action \eqref{S_Q.op}. 
The kernel of the dual action is thereby desired to have a minimal dimension,
and it also must be an ideal in the symmetry algebra. 
From the symmetry commutators \eqref{boussinesq.sys.symm.alg}, 
the only $1$-dimensional ideal is $\spn(P_1)$, 
while the $2$-dimensional ideals consist of $\{P_1,P_2\}$, $\{P_3,P_4\}$. 
One can find the kernel of the dual action $S_{Q}(P)$ for $Q=\sum_{i=1..6} c_i Q_i$ 
from Table~\ref{table:boussinesqsys.2ndsymmaction}. 
The cases having dimension at most $2$ consist of 
$\ker(S_Q)=\spn(P_5 -\tfrac{c_4}{c_5}P_4 -\tfrac{c_3}{c_5}P_3)$ if $c_5\neq 0$,
and $\ker(S_Q)=\spn(P_3,P_4)$ if $c_5=0$, $c_6\neq0$. 
Only the latter case is an ideal. 
Then the range of $S_{Q^v}(P^v)$ turns out to be contained in 
$\spn(Q_1,Q_2,Q_4 -\tfrac{c_4}{c_3} Q_3,Q_6+ \tfrac{c_3{}^2+c_4{}^2}{c_3 c_6} Q_3)$, 
which will be the maximal domain for the adjoint-symmetry bracket. 
This yields the result shown in Table~\ref{table:boussinesqsys.2ndadjsymmbracket}. 
\begin{table}[h!]
\caption{Boussinesq system: adjoint-symmetry  bracket
from symmetry action \eqref{symmaction2.adjsymm} with 
$Q=Q_6 +c_4 Q_4 + c_3 Q_3 +c_2 Q_2 +c_1 Q_1$, where 
$Q_{4'}= Q_6+ \tfrac{c_3{}^2+c_4{}^2}{c_3} Q_3$, 
$Q_{3'} = Q_4 -\tfrac{c_4}{c_3} Q_3$,
$Q_{2'}=Q_2$, $Q_{1'}=Q_1$}
\label{table:boussinesqsys.2ndadjsymmbracket}
\centering
\begin{tabular}{l||c|c|c|c}
& $Q_{1'}$
& $Q_{2'}$
& $Q_{3'}$
& $Q_{4'}$
\\
\hline
\hline  
$Q_{1'}$
& $0$
& $0$
& $0$
& $\tfrac{1}{3} Q_{1'}$
\\
\hline
$Q_{2'}$
& 
& $0$
& $0$
& $\tfrac{2}{3} Q_{1'} - \tfrac{1}{3} Q_{2'}$
\\
\hline
$Q_{3'}$
& 
& 
& $0$
& $0$
\\
\hline
$Q_{4'}$
&
&
&
& $0$
\\
\end{tabular}
\end{table}

For the second action \eqref{symmaction1.adjsymm},
one can show from Table~\ref{table:boussinesqsys.2ndsymmaction} 
that none of the cases in which the kernel of the dual action $S_{Q}(P)$ 
for $Q=\sum_{i=1..6} c_i Q_i$ 
is at most $2$-dimensional are ideals. 
The case having minimal dimension such that the kernel is an ideal 
turns out to be $\ker(S_Q) = \spn(P_1,P_2,P_3,P_4)$, which arises when $c_5=c_6=0$. 
The maximal domain of the resulting adjoint-symmetry bracket is then $2$ dimensional. 

Finally, the third symmetry action \eqref{symmaction1.adjsymm},
which is shown in Table~\ref{table:boussinesqsys.3rdsymmaction},
is non-trivial only on the non-multiplier $Q_6$. 
This action directly yields a  bracket on the whole space 
$\spn(Q_1,Q_2,Q_3,Q_4,Q_5,Q_6)$, 
which is isomorphic to symmetry algebra. 

\begin{table}[h!]
\caption{Boussinesq system: symmetry action \eqref{symmaction3.adjsymm} on the non-multiplier adjoint-symmetry}
\label{table:boussinesqsys.3rdsymmaction}
\centering
\begin{tabular}{l||c|c|c|c|c|c}
& $P_1$
& $P_2$
& $P_3$
& $P_4$
& $P_5$
& $P_6$
\\
\hline
\hline  
$Q_6$
& $3Q_1$
& $3Q_2$
& $3Q_3$
& $3Q_4$
& $3Q_5$
& $3Q_6$
\\
\end{tabular}
\end{table}

\subsection{Symplectic 2-form and Hamiltonian operator}
The third symmetry action encodes a Noether operator $\Jop$ 
(cf section~\ref{sec:noetherop}) 
for the Boussinesq system \eqref{boussinesq.sys}.
Specifically, one has 
\begin{equation}\label{boussinesq.sys.Jop}
\Jop = \tfrac{1}{3} (Q_6' +R^*_{Q_6}) = \begin{pmatrix} 0 & -1 \\ 1 & 0 \end{pmatrix} ,
\end{equation}
which is obtained from 
\begin{equation}
Q'_6 = \begin{pmatrix} 0 & -2 -t D_t & \\ 1 + t D_t \end{pmatrix},
\quad
R^*_{Q_6} = \begin{pmatrix} 0 & t D_t -1 \\ -t D_t +2 & 0 \end{pmatrix} .
\end{equation}
The factor $\tfrac{1}{3}$ has been inserted here as a convenient normalization for the sequel. 

There is a bilinear form \eqref{evol.2form} associated to the Noether operator \eqref{boussinesq.sys.Jop}:
\begin{equation}\label{boussinesq.sys.2form}
\w_{Q_6}(P,\tilde P) =  \int \tilde P \Jop(P)^\t \, dx 
= \int (\tilde P^h P^u - \tilde P^u P^h)\, dx ,
\end{equation}
where $P^u\partial_u +P^h\partial_h$ and $\tilde P^u\partial_u +\tilde P^h\partial_h$
are any pair of symmetries. 
This bilinear form is skew and closed (cf section~\ref{sec:2form}), 
whence it defines a symplectic 2-form. 
In particular, $\Jop$ is a symplectic operator. 

The inverse of $\Jop$ defines a Hamiltonian operator
\begin{equation}\label{boussinesq.sys.Hop}
\Hop = \Jop^{-1} = \begin{pmatrix} 0 & 1 \\ -1 & 0 \end{pmatrix} .
\end{equation}
This indicates that the Boussinesq system \eqref{boussinesq.sys} has a Hamiltonian formulation. 
If the time-derivative terms in each equation in the system 
are combined on one side, then 
\begin{equation}
(u_t - \Delta u_t,h_t -\Delta h_t)^\t 
= \Hop(\delta \mathcal{E}/\delta u,\delta \mathcal{E}/\delta h)^\t
\end{equation}
yields a Hamiltonian formulation 
in terms of the conserved energy, $\mathcal E$. 
This formulation can be expressed in the equivalent form 
\begin{equation}
\begin{pmatrix} u \\ h \end{pmatrix}_t
= (1-\Delta)^{-1}\Hop\begin{pmatrix} \delta \mathcal{E}/\delta u \\ \delta \mathcal{E}/\delta h\end{pmatrix}
\end{equation}
where $(1-\Delta)^{-1}\Hop$ is also a Hamiltonian operator. 

Thus, 
the symmetry action \eqref{evol.sys.S_P.3rd} in Table~\ref{table:boussinesqsys.3rdsymmaction}
involving the non-multiplier adjoint-symmetry 
directly encodes the Hamiltonian structure of the Boussinesq system \eqref{boussinesq.sys}. 
Moreover, because the Hamiltonian operator \eqref{boussinesq.sys.Hop} is algebraic, 
it directly yields a Lagrangian structure: 
\begin{equation}
\Hop \begin{pmatrix} G^u \\ G^h \end{pmatrix}
= \begin{pmatrix} E_u(L) \\ E_h(L) \end{pmatrix},
\quad
L = (h - \Delta h) u_t +\tfrac{1}{2}( h^2 + h(u_x^2 +u_y^2),
\end{equation}
where $G^u$ and $G^h$ denote the respective PDEs in the Boussinesq system \eqref{boussinesq.sys}. 
Here $E_u$ and $E_h$ are the Euler operators with respect to $u$ and $h$.

\section{Coupled solitary wave equations}\label{sec:coupledsolitarywave}

The near-resonant interaction of weakly nonlinear solitary waves 
can be described by a coupled system of KdV equations \cite{Gri2013}. 
Consider, in particular, the nonlinearly-coupled equations
\begin{equation}\label{kdv.sys}
u_t + (uv)_x +u_{xxx}  =0, 
\quad
v_t + uu_x + \kappa v_{xxx} =0, 
\end{equation}  
after scaling of the variables, 
where $\kappa\neq0$ is a relative dispersion parameter. 
Here $u(t,x)$ and $v(t,x)$ are the amplitudes of the two waves. 
This system models \cite{MajBie,BieMaj2004}
the energy transfer between Rossby waves 
in equatorial latitudes and mid latitudes in the atmosphere. 

Symmetries (in evolutionary form) $P^u\partial_u+P^v\partial_v$ 
are determined by the equations
\begin{subequations}\label{kdv.sys.symm.deteqns}
\begin{align}
( D_t P^u +  D_x (u P^v + v P^u) +D_x^3 P^u  )|_\Esp{G} =0, 
\\
( D_t P^v + D_x (u P^u) + \kappa D_x^3 P^v  )|_\Esp{G} =0, 
\end{align}
\end{subequations}  
where $\Esp{G}$ denotes the solution space of the system \eqref{kdv.sys}.
Adjoint-symmetries $(Q^u,Q^v)$ are determined by the adjoint equations
\begin{subequations}\label{kdv.sys.adjsymm.deteqns}
\begin{align}
( {-}D_t Q^u -v D_x Q^u - u D_x Q^v -D_x^3 Q^u  )|_\Esp{G} =0, 
\\
( {-}D_t Q^v - u D_x Q^u -\kappa D_x^3 Q^v  )|_\Esp{G} =0 . 
\end{align}
\end{subequations}  
In the general notation \eqref{pde.sys} for PDE systems, 
here $G=(G^u,G^v)$ denotes the two respective equations \eqref{kdv.sys}.

The linear space of Lie point symmetries of this system \eqref{kdv.sys}, 
with $P=(P^u,P^v)$, 
is generated by a time-translation, a space-translation, and a scaling:
\begin{equation}\label{kdv.sys.symms}
\begin{aligned}
&
P_1 = (u_t,v_t),
\quad
P_2= (u_x,v_x),
\quad
P_3 =(2u + x u_x  +3t u_t,2v +x v_x + 3t v_t) . 
\end{aligned}
\end{equation}
Their algebra has the non-zero commutators
\begin{equation}\label{kdv.sys.symm.alg}
[P_1,P_3] = -3 P_1, 
\quad
[P_2,P_3] = - P_2 . 
\end{equation}

The linear space of low-order adjoint-symmetries, $Q=(Q^u,Q^v)$, 
is given by the basis
\begin{equation}\label{kdv.sys.adjsymms}
\begin{aligned}
&
Q_1 = (1,0),
\quad
Q_2= (0,1),
\quad
Q_3 =(u,v), 
\quad
Q_4 =(uv +u_{xx},\tfrac{1}{2}u^2 +\kappa v_{xx}) . 
\end{aligned}
\end{equation}
These adjoint-symmetries are also multipliers for conservation laws of 
(up to an overall factor) 
the mass $\mathcal M^u = \int_\Rnum u\,dx$ and $\mathcal M^v = \int_\Rnum v\,dx$
for each wave, 
the combined momentum of the waves
$\mathcal P = \int_\Rnum (u^2 + v^2) \,dx$, 
and the net energy of the waves 
$\mathcal E = \int_\Rnum \tfrac{1}{2}(u_x^2 + \kappa v_x^2 -v u^2)\,dx$.

One sees that the third symmetry action \eqref{evol.symmaction3.adjsymm} is trivial
(cf section~\ref{sec:symmaction}), 
while the other two symmetry actions \eqref{evol.symmaction2.adjsymm} and \eqref{evol.symmaction1.adjsymm}
are given by the linear operator \eqref{evol.sys.S_P.1st}. 
This action is summarized in Table~\ref{table:kdvsys.1stsymmaction}. 
For evaluating the symmetry actions, 
all $t$-derivatives of $u$ and $v$ are replaced through equations \eqref{kdv.sys}. 

\begin{table}[h!]
\caption{Coupled solitary wave equations: symmetry action \eqref{evol.sys.S_P.1st} on adjoint-symmetries}
\label{table:kdvsys.1stsymmaction}
\centering
\begin{tabular}{l||c|c|c}
& $P_1$
& $P_2$
& $P_3$
\\
\hline
\hline  
$Q_1$
& $0$
& $0$
& $Q_1$
\\
\hline
$Q_2$
& $0$
& $0$
& $Q_2$
\\
\hline
$Q_3$
& $0$
& $0$
& $3 Q_3$
\\
\hline
$Q_4$
& $0$
& $0$
& $5 Q_4$
\\
\end{tabular}
\end{table}

\subsection{A nonlocal adjoint-symmetry and associated adjoint-symmetry brackets}
From the symmetry action shown in Table~\ref{table:kdvsys.1stsymmaction}, 
the dual action has $\ker S_Q(P)=\spn(P_1,P_2)$ for any choice of adjoint-symmetry $Q$. 
Hence, the cokernel, which is given by $\spn(P_3)$, is only $1$-dimensional. 
Since the dimension of the cokernel gives the dimension of the maximal domain 
on which a corresponding adjoint-symmetry  bracket can be defined, 
the resulting bracket is trivial. 

However, the situation changes when one considers the possibility of 
nonlocal adjoint-symmetries arising from potentials 
obtained via the mass conservation laws.
These potentials are given by $u=U_x$ and $v=V_x$, 
and they satisfy the coupled system
\begin{equation}\label{kdvpot.sys}
U_t + U_x V_x +U_{xxx}  =0, 
\quad
V_t + \tfrac{1}{2} U_x^2 + \kappa V_{xxx} =0 . 
\end{equation}  
Adjoint-symmetries $(Q^U,Q^V)$ of this system are determined by the equations
\begin{subequations}\label{kdvpot.sys.adjsymm.deteqns}
\begin{align}
( {-}D_t Q^U -D_x( V_x Q^U + U_x Q^V) -D_x^3 Q^U  )|_\Esp{G} =0, 
\\
( {-}D_t Q^V - D_x(U_x Q^U) -\kappa D_x^3 Q^V  )|_\Esp{G} =0 . 
\end{align}
\end{subequations}  
Note the relation 
\begin{equation}\label{kdv.sys.adjsymm.rel}
(Q^U,Q^V)=-D_x(Q^u,Q^v) 
\end{equation}
holds directly from the adjoint-symmetry equations \eqref{kdv.sys.adjsymm.deteqns} and \eqref{kdvpot.sys.adjsymm.deteqns}. 

The linear space of low-order adjoint-symmetries $(Q^U,Q^V)$ is generated by 
three adjoint-symmetries, 
two of which are inherited from the adjoint-symmetries $Q_3$ and $Q_4$ 
of the original system \eqref{kdv.sys} for $u,v$,
through the relation \eqref{kdv.sys.adjsymm.rel}. 
The other low-order adjoint-symmetry is given by 
\begin{equation}\label{kdvpot.scal.adjsymm}
(Q^U,Q^V) = -( 2U_x + xU_{xx}  +3t U_{tx}, 2V_x +xV_{xx} + 3t V_{tx} ) . 
\end{equation}
Applying the inverse relation 
\begin{equation}
(Q^u,Q^v)=-D_x^{-1}(Q^U,Q^V) , 
\end{equation}
one obtains the nonlocal adjoint-symmetry 
\begin{equation}\label{kdv.scal.adjsymm}
Q_5 =( U + xU_x  +3t U_t, V +xV_x + 3t V_t ) 
= \big( U + x u  -3t(u v +\kappa u_{xx}), V +x v - 3t(\tfrac{1}{2} u^2 + v_{xx}) \big) 
\end{equation}
admitted by the system \eqref{kdv.sys} for $u,v$. 
One can straightforwardly show that this adjoint-symmetry is not a multiplier. 

When the first symmetry action, as given by the linear operator \eqref{evol.sys.S_P.1st},
is applied to $Q_5$, one obtains
\begin{equation}\label{kdvsys.1stsymmaction.Qnonloc}
S_{P_1}(Q_5) = 5 Q_4,
\quad
S_{P_2}(Q_5) = -3 Q_3,
\quad
S_{P_3}(Q_5) = 0, 
\end{equation}
by using the variational derivative relations 
$E_u = -D_x^{-1} E_U$ and $E_v = -D_x^{-1} E_V$. 
Consequently, 
if one chooses 
\begin{equation}
Q=Q_5+c_2 Q_2 +c_1Q_1 := Q_{5'}
\end{equation}
with at least one of $c_1,c_2$ being non-zero, 
then $\ker S_Q(P)$ is empty,
and so the cokernel consists of the whole linear space of Lie point symmetries, 
$\spn(P_1,P_2,P_3)$. 
This choice of $Q$ produces a maximal domain 
for defining the adjoint-symmetry  bracket \eqref{adjsymm.bracket}. 

From equation \eqref{kdvsys.1stsymmaction.Qnonloc} and Table~\ref{table:kdvsys.1stsymmaction}, 
one obtains 
\begin{equation}\label{kdvsys.S_Q.nonloc}
\begin{aligned}
& 
S_{P_1}(Q_{5'}) = S_{Q_{5'}}(P_1) = 5Q_4, 
\quad
S_{P_2}(Q_{5'}) = S_{Q_{5'}}(P_2) = -3 Q_3 , 
\\& 
S_{P_3}(Q_{5'}) = S_{Q_{5'}}(P_3) = c_1 Q_1+c_2 Q_2 ,
\end{aligned}
\end{equation}
which yields 
\begin{equation}\label{kdvsys.inv.S_Q.nonloc}
S_{Q_{5'}}^{-1}(c_1 Q_1+c_2 Q_2) =  P_3, 
\quad
S_{Q_{5'}}^{-1}(Q_3) = -\tfrac{1}{3} P_2,
\quad
S_{Q_{5'}}^{-1}(Q_4) = \tfrac{1}{5} P_1 . 
\end{equation}
The resulting adjoint-symmetry  bracket \eqref{adjsymm.bracket}
on the linear subspace $\spn(c_1 Q_1 + c_2 Q_2,Q_3,Q_4)$
is shown in Table~\ref{table:kdvsys.pot.1stadjsymmbracket}. 
This bracket is a Lie bracket which is isomorphic to the symmetry algebra \eqref{kdv.sys.symm.alg}. 

\begin{table}[h!]
\caption{Coupled solitary wave equations:  adjoint-symmetry  bracket
from symmetry action \eqref{evol.sys.S_P.1st} with $Q=Q_5 +c_2 Q_2 +c_1Q_1$}
\label{table:kdvsys.pot.1stadjsymmbracket}
\centering
\begin{tabular}{c||c|c|c}
& $c_1Q_1+c_2 Q_2$
& $Q_3$
& $Q_4$
\\
\hline
\hline  
$c_1Q_1+c_2 Q_2$
& $0$
& $Q_3$
& $3Q_4$
\\
\hline
$Q_3$
& 
& $0$
& $0$
\\
\hline
$Q_4$
& 
& 
& $0$
\\
\end{tabular}
\end{table}

Since $Q_5$ is not a multiplier, 
the third symmetry action \eqref{evol.symmaction3.adjsymm} is now non-trivial. 
In terms of components $Q=(Q^u,Q^v)$ and $P=(P^u,P^v)$, 
the form of this symmetry action is given by the linear operator 
\begin{equation}\label{evol.sys.S_P.3rd}
S_P(Q) = \big( \pr\X_f(Q^u) - E_u(Q f^\t),\pr\X_f(Q^v) - E_v(Q f^\t) \big)\big|_{f=P}
\end{equation}
with $\X_f = f^u(t,x)\partial_u + f^v(t,x)\partial_v$, 
where $\pr\X$, $E_u$, $E_v$ are regarded as operators in total derivatives 
when $f=(f^u(t,x),f^v(t,x))$ is replaced by $P=(P^u,P^v)$. 
For $Q=Q_5$, 
the resulting action is summarized in Table~\ref{table:kdvsys.pot.3rdsymmaction}. 

\begin{table}[h!]
\caption{Coupled solitary wave equations: symmetry action \eqref{evol.sys.S_P.3rd} on the nonlocal adjoint-symmetry \eqref{kdv.scal.adjsymm}}
\label{table:kdvsys.pot.3rdsymmaction}
\centering
\begin{tabular}{l||c|c|c}
& $P_1$
& $P_2$
& $P_3$
\\
\hline
\hline  
$Q_5$
& $-2Q_4$
& $2Q_3$
& $2Q_5$
\\
\end{tabular}
\end{table}

The range of this action is the linear subspace of adjoint-symmetries 
$\spn(Q_3,Q_4,Q_5)$,
which provides a maximal domain for the adjoint-symmetry  bracket \eqref{adjsymm.bracket} 
with $Q=Q_5$,
as shown in Table~\ref{table:kdvsys.pot.3rdadjsymmbracket}. 
The resulting bracket is a Lie bracket which is isomorphic to the symmetry algebra \eqref{kdv.sys.symm.alg}. 

\begin{table}[h!]
\caption{Coupled solitary wave equations:  adjoint-symmetry  bracket from symmetry action \eqref{evol.sys.S_P.3rd} with $Q=Q_5$}
\label{table:kdvsys.pot.3rdadjsymmbracket}
\centering
\begin{tabular}{c||c|c|c}
& $Q_3$
& $Q_4$
& $Q_5$
\\
\hline
\hline  
$Q_3$
& $0$
& $0$
& $-\tfrac{1}{2}Q_3$
\\
\hline
$Q_4$
& 
& $0$
& $-\tfrac{3}{2}Q_4$
\\
\hline
$Q_5$
& 
& 
& $0$
\\
\end{tabular}
\end{table}

\subsection{Symplectic 2-form and Hamiltonian operator}
The symmetry action \eqref{evol.sys.S_P.3rd} 
constructed in terms of the nonlocal adjoint-symmetry \eqref{kdv.scal.adjsymm}
encodes a Noether operator $\Jop$ (cf section~\ref{sec:noetherop})
for the coupled KdV equations \eqref{kdv.sys}. 
Specifically, one has 
$\Jop(P) = S_P(Q_5) = \big( \pr\X_f Q_5^u - E_u(Q_5 f^\t),\pr\X_f Q_5^v - E_v(Q_5 f^\t) \big)\big|_{f=P}$, 
where
\begin{equation}
\begin{aligned}
& ( \pr\X_f Q_5^u, \pr\X_f Q_5^v )|_{f=P} \\
& = 
\big( D_x^{-1}P^u + x P^u  -3t(v P^u+u P^v +\kappa D_x^2 P^u), 
D_x^{-1} P^v +x P^v - 3t(u P^u + D_x^2 P^v) \big) 
\end{aligned}
\end{equation}
and
\begin{equation}
\begin{aligned}
& ( E_u(Q_5 f^\t), E_v(Q_5 f^\t) )|_{f=P} \\
& = 
\big( {-}D_x^{-1}P^u + x P^u  -3t(v P^u+u P^v +\kappa D_x^2 P^u), 
-D_x^{-1} P^v +x P^v - 3t(u P^u + D_x^2 P^v) \big) . 
\end{aligned}
\end{equation}
This yields, after scaling by a convenient normalization factor $\tfrac{1}{2}$, 
\begin{equation}\label{kdv.sys.Jop}
\Jop = \begin{pmatrix} D_x^{-1} & 0 \\ 0 & D_x^{-1} \end{pmatrix}
\end{equation}
which actually is a symplectic operator. 
In particular, there is a bilinear form \eqref{evol.2form} 
associated to this operator, 
\begin{equation}\label{kdv.sys.2form}
\w_{Q_5}(P,\tilde P) =  \int \tilde P \Jop(P)^\t \, dx 
= \int (\tilde P^u D_x^{-1} P^u + \tilde P^v D_x^{-1} P^v)\, dx ,
\end{equation}
where $P^u\partial_u +P^v\partial_v$ and $\tilde P^u\partial_u +\tilde P^v\partial_v$
are any pair of symmetries,
and $\t$ denotes the transpose. 
Modulo boundary terms, 
this bilinear form is skew and closed (cf section~\ref{sec:2form}), 
and hence it defines a symplectic 2-form. 

The inverse of $\Jop$ defines a Hamiltonian operator
\begin{equation}
\Hop = \Jop^{-1} = 
\begin{pmatrix} D_x & 0 \\ 0 &  D_x \end{pmatrix} .
\end{equation}
As a consequence, 
the coupled KdV equations \eqref{kdv.sys} have the Hamiltonian formulation
\begin{equation}
(u_t,v_t)^\t = \Hop(\delta \mathcal{E}/\delta u,\delta \mathcal{E}/\delta v)^\t
\end{equation}
in terms of the conserved energy, $\mathcal E$. 
From this formulation, one has 
$u_t = D_x(\delta \mathcal{E}/\delta u)$
and 
$v_t = D_x(\delta \mathcal{E}/\delta v)$,
which are analogous to the well-known first Hamiltonian structure of the KdV equation. 

The symmetry action \eqref{evol.sys.S_P.3rd} involving the nonlocal adjoint-symmetry \eqref{kdv.scal.adjsymm}
thereby directly encodes the Hamiltonian structure of 
the coupled KdV equations \eqref{kdv.sys}.

\section{Nonlinear acoustic equation}\label{sec:acoustic}

Nonlinear and dissipative effects in the propagation of sound waves 
in a compressible medium (like gases, liquids, or human tissue) \cite{HamBla-book}
can be modelled by the wave equation 
$p_{tt} - \beta (p^2)_{tt} - \alpha p_{ttt} = \Delta p$,
known as Westervelt's equation \cite{Wes}, 
where $\alpha>0$ is the damping coefficient
and $\beta>0$ is the nonlinearity coefficient which arises from the equation of state. 
Here $p(t,x,y,z)$ is the pressure fluctuation,
and units have been chosen so that the sound speed for small amplitude waves 
(i.e. the linear approximation) is $c=1$. 

For the situation of spherical waves, 
Westervelt's equation becomes
\begin{equation}\label{acoustic.eqn}
p_{tt} - \beta (p^2)_{tt} -\alpha p_{ttt} = p_{rr}+\tfrac{2}{r}p_r
\end{equation}
for $p(t,r)$,  where $r$ is radial variable. 

The determining equation for symmetries (in evolutionary form) 
$P\partial_p$ is given by 
\begin{equation}\label{acoustic.symm.deteqns}
( D_t^2 (P +2\beta p P) -\alpha D_t^3 P  -  D_r^2 P +\tfrac{2}{r} D_r P  )|_\Esp{G} =0,
\end{equation}  
where $\Esp{G}$ denotes the solution space of equation \eqref{acoustic.eqn}.
Lie point symmetries are comprised by 
a time-translation and two generalized scalings:
\begin{equation}\label{acoustic.symms1}
P_1 = p_t , 
\quad
P_2 =2p + 2t p_t +3r p_r -\tfrac{1}{\beta} ,
\end{equation}
and, when $\alpha=0$,
\begin{equation}\label{acoustic.symms2}
P_3 =t p_t + r p_r .
\end{equation}
The non-zero commutators in the symmetry algebra consist of 
\begin{equation}\label{acoustic.symm.alg}
[P_1,P_2] = -2 P_1 ,
\quad
[P_1,P_3] = - P_1 .
\end{equation}

Adjoint-symmetries $Q$ are determined by the adjoint of equation \eqref{acoustic.symm.deteqns}:
\begin{equation}\label{acoustic.adjsymm.deteqns}
( D_t^2 Q +2\beta p D_t^2 Q -  D_r^2 Q -\tfrac{2}{r} D_r Q  +\tfrac{2}{r^2} Q )|_\Esp{G} =0 . 
\end{equation}  
The linear space of low-order adjoint-symmetries is given by the basis
\begin{equation}\label{acoustic.adjsymms}
Q_1 = r,
\quad
Q_2= r^2
\quad
Q_3 =t\,r, 
\quad
Q_4 =t\,r^2 . 
\end{equation}
These adjoint-symmetries are also multipliers for conservation laws of 
(up to an overall factor) 
transverse momentum 
\begin{equation}
\mathcal P = \int_0^\infty ( (1-2\beta p)p_t -\alpha p_{tt} )\, r^2\,dr,
\end{equation}
center of transverse-momentum motion
\begin{equation}
\mathcal X = \int_0^\infty ( t(1-2\beta p)p_t -t\alpha p_{tt} +\alpha p_t +\beta p^2 -p )\, r^2\,dr
= t \mathcal P +\int_0^\infty ( \alpha p_t +\beta p^2 -p )\, r^2\,dr,
\end{equation}
and their radially-weighted counterparts 
\begin{equation}
\mathcal P^r = \int_0^\infty ( (1-2\beta p)p_t -\alpha p_{tt} )\, r\,dr,
\quad
\mathcal X^r = t {\mathcal P}^r +\int_0^\infty ( \alpha p_t +\beta p^2 -p )\, r\,dr,
\end{equation}
where the weighting factor is $\tfrac{1}{r}$. 

Because the symmetry space is two dimensional if $\alpha\neq0$, 
the maximal domain on which an adjoint-symmetry bracket can be defined 
is a $2$-dimensional subspace of the four-dimensional space of adjoint-symmetries. 
In the case $\alpha=0$, the maximal domain is at most $3$-dimensional. 

However, an adjoint-symmetry bracket on a much larger space can be found by 
considering the adjoint-symmetries and symmetries in a potential system. 
Potentials can be introduced in several different ways through each of the four conservation laws. 
The most useful potential arises through two layers as follows, 
using the conservation law for transverse momentum. 

First, in the standard way \cite{BCA-book}, 
put 
\begin{subequations}\label{acoustic.pot1sys}
\begin{align}
(1-2\beta p)p_t -\alpha p_{tt} ) & = r^{-2} u_r, 
\label{acoustic.pot1eqn1}
\\
p_r & = r^{-2} u_t , 
\label{acoustic.pot1eqn2}
\end{align}
\end{subequations}
which turns the wave equation \eqref{acoustic.eqn} into an identity
satisfied by the potential $u(t,r)$. 
Next, introduce a further potential $v(t,r)$ through equation \eqref{acoustic.pot1eqn2},
by putting
\begin{equation}\label{acoustic.pot2eqns}
p =  r^{-2} v_t, 
\quad
r^{-2} u = (r^{-2} v)_r .
\end{equation}
Then equation \eqref{acoustic.pot1eqn1} yields 
$v_{tt} -2\beta r^{-2} v_t v_{tt} -\alpha v_{ttt} = u_r$,
which gives a second layer potential system
\begin{equation}\label{acoustic.potsys}
v_r = 2r^{-1} v +u,
\quad
(v_t -\beta r^{-2} v_t^2 -\alpha v_{tt})_t = u_r , 
\end{equation}
or equivalently 
\begin{equation}\label{acoustic.pot2eqn}
(v_t -\beta r^{-2} v_t^2 -\alpha v_{tt})_t = (v_r -\tfrac{2}{r} v)_r . 
\end{equation}

\subsection{Adjoint-symmetry brackets arising from a potential system}
Symmetries (in evolutionary form) $P^v\partial_v$ of the potential equation \eqref{acoustic.pot2eqn}
are determined by the equation
\begin{equation}\label{acoustic.pot2eqn.symm.deteqns}
( D_t^2 P^v -2\beta r^{-2} D_t(v_t D_t P^v) -\alpha D_t^3 P^v  -D_r^2 P^v+ 2D_r(r^{-1} P^v) )|_\Esp{G^v} =0, 
\end{equation}  
where $\Esp{G^v}$ denotes the solution space of equation \eqref{acoustic.pot2eqn}.
Adjoint-symmetries $Q^v$ are determined by the adjoint equation
\begin{equation}\label{acoustic.pot2eqn.adjsymm.deteqns}
( D_t^2 Q^v -2\beta r^{-2} D_t(v_t D_t Q^v) +\alpha D_t^3 Q^v  -D_r^2 Q^v - 2r^{-1} D_r Q^v) )|_\Esp{G^v} =0 . 
\end{equation}  

The linear space of Lie point symmetries of the potential equation \eqref{acoustic.pot2eqn}
is generated by two shifts, 
in addition to the three point symmetries that are inherited 
from the acoustic wave equation \eqref{acoustic.eqn} 
via the relation 
\begin{equation}\label{acoustic.P.rel}
P=r^{-2}D_t P^v . 
\end{equation}
In particular, these symmetries are given by 
\begin{equation}\label{acoustic.pot2eqn.symms1}
P^v_1 = r^2,
\quad
P^v_2= r, 
\quad
P^v_3= v_t, 
\quad
P^v_4 =2 t v_t + 3 r v_r - 6 v - \tfrac{1}{\beta}t\, r^2 ,
\end{equation}
and, when $\alpha=0$, 
\begin{equation}\label{acoustic.pot2eqn.symms2}
P^v_5 =t v_t + r v_r - 3 v. 
\end{equation}
Their algebra has the non-zero commutators
\begin{equation}\label{acoustic.pot2eqn.symm.alg}
\begin{aligned}
& [P^v_1,P^v_5] = - P^v_1, 
\quad
[P^v_2,P^v_4] = -3 P^v_2, 
\quad
[P_2^v,P_5^v] = -2 P_2^v, 
\\
& [P_3^v,P_4^v] = \tfrac{1}{\beta} P_1^v -2 P_3^v , 
\quad
[P^v_3,P^v_5] = - P^v_3 .
\end{aligned}  
\end{equation}
Note that the shifts $P^v_1$ and $P^v_2$ project onto trivial symmetries $P=0$ 
of the acoustic wave equation \eqref{acoustic.eqn}. 

The linear space of low-order adjoint-symmetries 
is given by the basis
\begin{equation}\label{acoustic.pot2eqn.multrs1}
Q^v_1 = 1,
\quad
Q^v_2= r^{-1} , 
\end{equation}
and, when $\alpha =0$, 
\begin{subequations}\label{acoustic.pot2eqn.adjsymms2}
\begin{align}
&
Q^v_3 =r^{-2} v_t, 
\quad
Q^v_4 =r^{-2}(3 t v_t - 9 v) +7 r^{-1} v_r  -4\tfrac{1}{\beta} t, 
\label{acoustic.pot2eqn.multrs2}
\\& 
Q^v_5 = r^{-1} v_r - \tfrac{1}{\beta} t .
\label{acoustic.pot2eqn.nonmultr}
\end{align}
\end{subequations}
One can easily check that the adjoint symmetry \eqref{acoustic.pot2eqn.nonmultr} 
is not a conservation law multiplier. 
The two adjoint-symmetries \eqref{acoustic.pot2eqn.multrs2} 
are multipliers for conservation of 
energy **** check beta terms
\begin{equation}\label{acoustic.ener}
\mathcal E = \int_0^\infty ( \tfrac{1}{2} p^2 -\tfrac{2}{3}\beta p^3  +\tfrac{1}{2} r^{-4} v_r^2 - r^{-6} v^2 )\, r^2\,dr,
\end{equation}
and a dilation energy 
\begin{equation}
\mathcal W = \int_0^\infty \big( t(\tfrac{11}{2} p^2 -2\beta p^3 -4\tfrac{1}{\beta} p +\tfrac{3}{2}r^{-2} v_r^2 -3 r^{-6} v^2) +7 r^{-1}(p -\beta p^2)  -(9p -9\beta p^2 -4\tfrac{1}{\beta})v \big)\, r^2\,dr,
\end{equation}
respectively. 
Finally, the first two adjoint-symmetries \eqref{acoustic.pot2eqn.multrs1} 
are inherited from $Q_3$ and $Q_4$ via the relation 
\begin{equation}\label{acoustic.Q.rel}
Q^v =-r^{-2}D_t Q
\end{equation} 
which follows directly from the adjoint-symmetry determining equations
\eqref{acoustic.adjsymm.deteqns} and \eqref{acoustic.pot2eqn.adjsymm.deteqns}. 
These inherited adjoint-symmetries are thus 
the respective multipliers for conservation of $\mathcal X$ and $\mathcal X^r$. 
(Note that $\mathcal P$ and $\mathcal P^r$ are locally trivially 
when they are evaluated on $\Esp{G^v}$ in terms of the potential $v$.)

The three actions of the symmetries \eqref{acoustic.pot2eqn.symms1}--\eqref{acoustic.pot2eqn.symms2}
on the adjoint-symmetries \eqref{acoustic.pot2eqn.multrs1}--\eqref{acoustic.pot2eqn.adjsymms2}
give rise to adjoint-symmetry brackets at the level of the potential equation \eqref{acoustic.pot2eqn}. 
This structure will not be preserved under projection \eqref{acoustic.P.rel} and \eqref{acoustic.Q.rel}
back to the acoustic wave equation, 
since two of the potential symmetries are lost,
in contrast to the example of the coupled solitary wave equations 
considered in the previous section. 

The first and second actions \eqref{symmaction2.adjsymm}, \eqref{symmaction1.adjsymm}
here differ only on the non-multiplier $Q_5$. 
They are computed from the symmetry and adjoint-symmetry expressions by use of: 
\begin{equation}
R_{P^v_1} = R_{P^v_2} = 0, 
\quad
R_{P^v_3} = D_t , 
\quad
R_{P^v_4} = 2t D_t + 3r D_r , 
\quad
R_{P^v_5} = t D_t + r D_r - 1 ,
\end{equation}
and 
\begin{equation}
R_{Q^v_1} =R_{Q^v_2} = 0, 
\quad
R_{Q^v_3} = r^-2 D_t, 
\quad
R_{Q^v_4} = r^{-2}( 3t D_t + 7r D_r + 5 ), 
\quad
R_{Q^v_5} = r^{-2} (r D_r + 2) .
\end{equation}
This leads to the results shown in Tables~\ref{table:acoustic.potsys.2ndsymmaction} and~\ref{table:acoustic.potsys.1stsymmaction}. 

\begin{table}[h!]
\caption{Acoustic potential equation: symmetry action \eqref{symmaction2.adjsymm} on adjoint-symmetries}
\label{table:acoustic.potsys.2ndsymmaction}
\centering
\begin{tabular}{l||c|c|c|c|c}
& $P^v_1$
& $P^v_2$
& $P^v_3$
& $P^v_4$
& $P^v_5$
\\
\hline
\hline  
$Q^v_1$
& $0$
& $0$
& $0$
& $-5Q^v_1$
& $-3Q^v_1$
\\
\hline
$Q^v_2$
& $0$
& $0$
& $0$
& $-2Q^v_2$
& $-2Q^v_2$
\\
\hline
$Q^v_3$
& $0$
& $0$
& $0$
& $-\tfrac{1}{\beta}Q^v_1 -3Q^v_3$
& $-3Q^v_3$
\\
\hline
$Q^v_4$
& $5Q^v_1$
& $-2Q^v_2$
& $4\tfrac{1}{\beta}Q^v_1 -3Q^v_3$
& $-5Q^v_4$
& $-4Q^v_4$
\\
\hline
$Q^v_5$
& $2Q^v_1$
& $Q^v_2$
& $\tfrac{1}{\beta} Q^v_1$
& $-5Q^v_5$
& $-4Q^v_5$
\\
\end{tabular}
\end{table}

\begin{table}[h!]
\caption{Acoustic potential equation: symmetry action \eqref{symmaction1.adjsymm} on adjoint-symmetries}
\label{table:acoustic.potsys.1stsymmaction}
\centering
\begin{tabular}{l||c|c|c|c|c}
& $P^v_1$
& $P^v_2$
& $P^v_3$
& $P^v_4$
& $P^v_5$
\\
\hline
\hline  
$Q^v_1$
& $0$
& $0$
& $0$
& $-5Q^v_1$
& $-3Q^v_1$
\\
\hline
$Q^v_2$
& $0$
& $0$
& $0$
& $-2Q^v_2$
& $-2Q^v_2$
\\
\hline
$Q^v_3$
& $0$
& $0$
& $0$
& $-\tfrac{1}{\beta} Q^v_1 -3Q^v_3$
& $-3Q^v_3$
\\
\hline
$Q^v_4$
& $5Q^v_1$
& $-2Q^v_2$
& $4\tfrac{1}{\beta} Q^v_1 -3Q^v_3$
& $-5Q^v_4$
& $-4Q^v_4$
\\
\hline
$Q^v_5$
& $-Q^v_1$
& $-2Q^v_2$
& $\tfrac{1}{\beta} Q^v_1 -3Q^v_3$
& $-2Q^v_4$
& $-Q^v_4$
\\
\end{tabular}
\end{table}

The third action \eqref{symmaction3.adjsymm}, 
which is shown in Table~\ref{table:acoustic.potsys.3rdsymmaction},
is non-trivial only on the non-multiplier $Q_5$. 

\begin{table}[h!]
\caption{Acoustic potential equation: symmetry action \eqref{symmaction3.adjsymm} on the non-multiplier adjoint-symmetry}
\label{table:acoustic.potsys.3rdsymmaction}
\centering
\begin{tabular}{l||c|c|c|c|c}
& $P^v_1$
& $P^v_2$
& $P^v_3$
& $P^v_4$
& $P^v_5$
\\
\hline
\hline  
$Q^v_5$
& $3Q^v_1$
& $3Q^v_2$
& $3Q^v_3$
& $2Q^v_4 -5Q^v_5$
& $Q^v_4 -4Q^v_5$
\\
\end{tabular}
\end{table}

Now, the adjoint-symmetry brackets arising from the dual actions \eqref{S_Q.op}
will be considered. 
The maximal domain for defining a bracket is given by having a maximal range for the dual action and, therefore, a minimal dimension for the kernel 
which also must be an ideal in the symmetry algebra. 
From the symmetry commutators \eqref{acoustic.pot2eqn.symm.alg}, 
the ideals with smallest dimension consist of $\spn(P^v_1)$, $\spn(P^v_2)$, 
which are $1$-dimensional.

For the first symmetry action, 
shown in Table~\ref{table:acoustic.potsys.2ndsymmaction},
the dual action $S_{Q^v}(P^v)$ given by $Q^v=\sum_{i=1..5} c_i Q^v_i$ 
will have $\spn(P^v_1)$ as the kernel if $c_5=1$, $c_4= -\tfrac{2}{5}$, 
and $\spn(P^v_2)$ as the kernel if $c_5=1$, $c_4= \tfrac{1}{2}$. 
In both cases, the range of $S_{Q^v}(P^v)$ is contained in 
$\spn(Q^v_1,Q^v_2,Q^v_3,Q^v_5 +c_4 Q^v_4)$, 
which will be the maximal domain for the resulting adjoint-symmetry bracket. 
This yields the  brackets shown in Tables~\ref{table:acoustic.potsys.2ndadjsymmbracket1} and~\ref{table:acoustic.potsys.2ndadjsymmbracket2}. 

\begin{table}[h!]
\caption{Acoustic potential equation: adjoint-symmetry  bracket
from symmetry action \eqref{symmaction2.adjsymm} with 
$Q^v=Q^v_5 -\tfrac{2}{5} Q^v_4 +c_3 Q^v_3+c_2 Q^v_2 +c_1 Q^v_1$, where 
$Q^v_{4'}= \lambda(Q^v_5 -\tfrac{2}{5} Q^v_4)$, 
$Q^v_{3'}=\lambda Q^v_3$, $Q^v_{2'}=\lambda Q^v_2$,  $Q^v_{1'} = \lambda(Q^v_1 -2\beta Q^v_3)$,
$\lambda = c_3 +2\beta c_1$} 
\label{table:acoustic.potsys.2ndadjsymmbracket1}
\centering
\begin{tabular}{l||c|c|c|c}
& $Q^v_{1'}$
& $Q^v_{2'}$
& $Q^v_{3'}$
& $Q^v_{4'}$
\\
\hline
\hline  
$Q^v_{1'}$
& $0$
& $-\tfrac{4\beta}{5} Q^v_{2'}$
& $\tfrac{3}{5} Q^v_{1'}$
& $-\tfrac{3c_3}{5} Q^v_{1'} + \tfrac{4\beta c_2}{5} Q^v_{2'}$
\\
\hline
$Q^v_{2'}$
& 
& $0$
& $\tfrac{2}{5} Q^v_{2'}$
& $\tfrac{c_3+2\beta c_1}{5} Q^v_{2'}$
\\
\hline
$Q^v_{3'}$
& 
& 
& $0$
& $\tfrac{\beta c_1-c_3}{5\beta} Q^v_{1'} + \tfrac{2c_2}{5} Q^v_{2'}$
\\
\hline
$Q^v_{4'}$
&
&
&
& $0$
\\
\end{tabular}
\end{table}

\begin{table}[h!]
\caption{Acoustic potential equation: adjoint-symmetry  bracket
from symmetry action \eqref{symmaction2.adjsymm} with 
$Q^v=Q^v_5 +\tfrac{1}{2} Q^v_4 +c_3 Q^v_3+ c_2 Q^v_2 +c_1 Q^v_1$, where 
$Q^v_{4'}= Q^v_5 +\tfrac{1}{2} Q^v_4$, 
$Q^v_{3'}= Q^v_3$, $Q^v_{2'}= c_2 Q^v_2$,  $Q^v_{1'} = Q^v_1$}
\label{table:acoustic.potsys.2ndadjsymmbracket2}
\centering
\begin{tabular}{l||c|c|c|c}
& $Q^v_{1'}$
& $Q^v_{2'}$
& $Q^v_{3'}$
& $Q^v_{4'}$
\\
\hline
\hline  
$Q^v_{1'}$
& $0$
& $\tfrac{5}{2} Q^v_{1'}$
& $0$
& $-Q^v_{1'}$
\\
\hline
$Q^v_{2'}$
& 
& $0$
& $-\tfrac{2}{\beta} Q^v_{1'} + \tfrac{3}{2} Q^v_{3'}$
& $\tfrac{4c_3+5\beta c_1}{2\beta} Q^v_{1'} -\tfrac{3c_3}{2} Q^v_{3'}$
\\
\hline
$Q^v_{3'}$
& 
& 
& $0$
& $\tfrac{1}{\beta} Q^v_{1'} + Q^v_{3'}$
\\
\hline 
$Q^v_{4'}$
&
&
&
& $0$
\\
\end{tabular}
\end{table}

For the second symmetry action, 
shown in Table~\ref{table:acoustic.potsys.1stsymmaction},
the dual action $S_{Q^v}(P^v)$ given by $Q^v=\sum_{i=1..5} c_i Q^v_i$ 
never has $\spn(P^v_1)$ or $\spn(P^v_2)$ as the kernel. 
In particular, 
in the first case, one needs $c_5=1$, $c_4= \tfrac{1}{5}$, 
but then the kernel also contains 
$\spn(P^v_5 -\tfrac{3}{5} P^v_4 -\tfrac{c_3}{3}P^v_3 -\tfrac{c_2}{3}P^v_2)$; 
and in the second case, one needs $c_5=1$, $c_4= -1$, 
whereby the kernel then also contains
$\spn(P^v_3 -\tfrac{1}{2\beta}P^v_1,P^v_5 -P^v_4+\tfrac{c_3+2\beta c_1}{6\beta}P^v_1)$. 
In both cases, the kernel is at least $2$-dimensional, 
and hence the maximal domain on which an adjoint-symmetry bracket can be defined
is at most $3$-dimensional,
which is no larger than the domain arising at the level of the acoustic wave equation. 

Finally, 
for the third symmetry action, 
which appears in Table~\ref{table:acoustic.potsys.3rdsymmaction}, 
the dual action with $Q^v=Q^v_5$ has an empty kernel, 
whence an adjoint-symmetry bracket is obtained 
on the whole space of adjoint-symmetries $\spn(Q^v_1,Q^v_2,Q^v_3,Q^v_4,Q^v_5)$. 
This bracket is shown in Table~\ref{table:acoustic.potsys.3rdadjsymmbracket}
and one sees that it is isomorphic to symmetry algebra
through the correspondence
$Q^v_1 \leftrightarrow P^v_1$, 
$Q^v_2 \leftrightarrow P^v_2$, 
$Q^v_3 \leftrightarrow P^v_3$, 
$Q^v_4 \leftrightarrow 4 P^v_4 - 5 P^v_5$, 
$Q^v_5 \leftrightarrow P^v_4 - 2 P^v_5 $. 

\begin{table}[h!]
\caption{Acoustic potential equation: adjoint-symmetry  bracket
from symmetry action \eqref{symmaction3.adjsymm} with $Q^v=Q^v_5$}
\label{table:acoustic.potsys.3rdadjsymmbracket}
\centering
\begin{tabular}{l||c|c|c|c|c}
& $Q^v_1$
& $Q^v_2$
& $Q^v_3$
& $Q^v_4$
& $Q^v_5$
\\
\hline
\hline  
$Q^v_1$
& $0$
& $0$
& $0$
& $\tfrac{5}{3} Q^v_1$
& $\tfrac{2}{3} Q^v_1$
\\
\hline
$Q^v_2$
& 
& $0$
& $0$
& $-\tfrac{2}{3} Q^v_2$
& $\tfrac{1}{3} Q^v_2$
\\
\hline
$Q^v_3$
& 
& 
& $0$
& $\tfrac{4}{3\beta} Q^v_1 -Q^v_3$
& $\tfrac{1}{3\beta} Q^v_1$
\\
\hline
$Q^v_4$
& 
&
&
& $0$
& $0$
\\
\hline
$Q^v_5$
&
&
&
&
& $0$
\\
\end{tabular}
\end{table}

\subsection{Noether operator and variational structure}
The third symmetry action \eqref{symmaction3.adjsymm} 
encodes a Noether operator $\Jop$ 
for the acoustic potential equation \eqref{acoustic.potsys} with $\alpha=0$,
namely when there is no damping. 

This operator is given by
$\Jop^v = Q^v_5{}' +R^*_{Q^v_5} = 3 r^{-2}$, 
since one has 
\begin{equation}
Q^v_5{}' = r^{-1} D_r, 
\quad
R^*_{Q^v_5} = r^{-2}(3 - rD_r)  .
\end{equation}
For convenience, a factor $\tfrac{1}{3}$ will be inserted in hereafter so that 
\begin{equation}\label{acoustic.potsys.Jop}
\Jop^v = r^{-2} .
\end{equation}

The Noether operator \eqref{acoustic.potsys.Jop} 
provides a mapping of symmetries into adjoint-symmetries:
\begin{equation}
\Jop^v(P^v) = r^{-2} P^v = Q^v .
\end{equation}
Specifically, in terms of the respective symmetry basis \eqref{acoustic.pot2eqn.symms1}--\eqref{acoustic.pot2eqn.symms2} 
and adjoint-symmetry basis \eqref{acoustic.pot2eqn.multrs1}--\eqref{acoustic.pot2eqn.adjsymms2}, 
one sees that 
$\Jop^v(P^v_1)=Q^v_1$, $\Jop^v(P^v_2)=Q^v_2$, $\Jop^v(P^v_3)=Q^v_3$, 
$\Jop^v(P^v_4)=\tfrac{2}{3}Q^v_4 -\tfrac{5}{3}Q^v_5$, 
$\Jop^v(P^v_5)=\tfrac{1}{3}Q^v_4 -\tfrac{4}{3}Q^v_5$.

As the Noether operator is algebraic, 
it yields a Lagrangian structure: 
\begin{equation}
\Jop^v(G^v) = E_v(L)
\end{equation}
where $G^v$ denotes the potential equation \eqref{acoustic.potsys} with $\alpha =0$,
and where the Lagrangian is straightforwardly found to be 
\begin{equation}\label{acoustic.potsys.Lagr}
L = \tfrac{1}{2} r^{-2}(v_r^2 -v_t^2)  +r^{-4}(\tfrac{1}{3}\beta v_t^3 - v^2) . 
\end{equation}
Here $E_v$ is the Euler operator with respect to $v$.

The preceding structure can be lifted to the acoustic wave equation \eqref{acoustic.eqn} 
through the relations \eqref{acoustic.Q.rel} and 
\begin{equation}
r^{-2}D_t G^v = G
\end{equation}
where $G$ denotes the PDE \eqref{acoustic.eqn} with $\alpha=0$. 
These relations imply that 
\begin{equation}
\Jop = -r^2 D_t^{-1} \Jop^v r^2 D_t^{-1} = -r^2 D_t^{-2}
\end{equation}
defines a Noether operator for the acoustic wave equation \eqref{acoustic.eqn} 
with no damping, $\alpha=0$. 
In particular, using the form of the potential $p =  r^{-2} v_t$, 
one has
\begin{subequations}
\begin{align}
\Jop(P_1) & =-r^2 D_t^{-2}(p_t) = -v , 
\\ 
\Jop(P_2) & =-r^2 D_t^{-2}(2p + 2t p_t +3r p_r -\tfrac{1}{\beta})
= \tfrac{1}{\beta} t^2 r^2 -2 t v_t - 3r \partial_t^{-1}v_r +8 \partial_t^{-1}v , 
\\
\Jop(P_3) & =-r^2 D_t^{-2}(t p_t +r p_r)
= v - \partial_t^{-1}(t v)  -r\partial_t^{-1}v_r  + 2\partial_t^{-1}v , 
\end{align}
\end{subequations}
all of which can be verified to be nonlocal adjoint-symmetries $Q$. 

The Lagrangian structure becomes 
\begin{equation}
\Jop(G) = E_p(L)
\end{equation}
through the variational derivative relation $\delta/\delta v = -r^{-2}D_t \delta/\delta p$,
where $L$ is a nonlocal expression \eqref{acoustic.potsys.Lagr} in terms of $p$:
\begin{equation}
L = r^2 \big( \tfrac{1}{2} (\partial_t^{-1} p_r)^2 -\tfrac{1}{2} p^2 + \tfrac{1}{3}\beta p^3 \big) -(\partial_t^{-1} p)^2  . 
\end{equation}

Thus, 
the symmetry action in Table~\ref{table:acoustic.potsys.3rdsymmaction}
involving the non-multiplier adjoint-symmetry 
directly encodes a variational structure for the undamped acoustic wave equation. 

\subsection{Hamiltonian structure}
The variational structure can also be lifted to the level of 
the first-layer potential system \eqref{acoustic.pot1sys}. 
Denote the PDEs in this system by
$G^1= p_r - r^{-2} u_t$ and $G^2=w_r - r^2(p-\beta p^2)_t$,
which satisfy the relation 
\begin{equation}\label{acoustic.pot1sys.rel}
D_r (r^2 G^1) + D_t G^2 = -D_t G^v = -r^2 G . 
\end{equation}

The determining equations for an adjoint-symmetry $(Q^1,Q^2)$ of 
the potential system $(G^1,G^2)=0$ are given by 
\begin{equation}
( -D_x Q^1 + r^2 (1-2\beta p) D_t Q^2 )|_\Esp{(G^1,G^2)} =0,
\quad
( -D_x Q^2 + r^{-2} D_t Q^1 )|_\Esp{(G^1,G^2)} =0 .
\end{equation}
From the PDE relation \eqref{acoustic.pot1sys.rel}, one can show that
\begin{equation}
Q^1 = -r^2 D_r D_t^{-1} Q^v = r^2 D_r (r^{-2} Q),
\quad
Q^2 = -Q^v = r^2 D_t Q .
\end{equation}
Similarly, a symmetry $P\partial_p + P^u\partial_u$ of the potential system satisfies
\begin{equation}
P  = r^{-2} D_t P^v, 
\quad
P^u = r^2 D_r (r^{-2} P^v) .
\end{equation}

Now consider the non-multiplier adjoint-symmetry \eqref{acoustic.pot2eqn.nonmultr}. 
The corresponding adjoint-symmetry $(Q^1,Q^2)$ of the potential system $(G^1,G^2)=0$ 
is given by 
\begin{equation}
\begin{aligned}
Q^1 & = -r^2 D_r D_t^{-1}(r^{-1} v_r - \tfrac{1}{\beta} t )
= -(\partial_t^{-1} u + r^3 (p-\beta p^2)), 
\\
Q^2 & = -(r^{-1} v_r - \tfrac{1}{\beta} t ) 
= -(r^{-1} u + 2 \partial_t^{-1} p) + \tfrac{1}{\beta} t, 
\end{aligned}
\end{equation}
with the use of equations \eqref{acoustic.pot1sys} and \eqref{acoustic.pot2eqns}
for the potentials. 
A straightforward computation then yields
\begin{equation}
Q^1{}'(P,P^u) = -(D_t^{-1} P^u +r^3(1-2\beta p)P), 
\quad
Q^2{}'(P,P^u) = -(r^{-1} P^u + 2 D_t^{-1} P) ,
\end{equation}
and 
\begin{equation}
R_{Q^1}^*(P,P^u) = r^3 (1-2\beta p) P, 
\quad
R_{Q^2}^*(P,P^u) = D_t^{-1} P +r^{-1} P^u .
\end{equation}
Hence, one obtains
\begin{equation}
\begin{pmatrix} 
Q^1{}'(P,P^u) +R_{Q^1}^*(P,P^u) \\ Q^2{}'(P,P^u) +R_{Q^2}^*(P,P^u) 
\end{pmatrix}
= \begin{pmatrix} -D_t^{-1} P^u  \\ -D_t^{-1} P \end{pmatrix} 
= \Jop\begin{pmatrix} P \\ P^u \end{pmatrix} ,
\end{equation}
giving the Noether operator 
\begin{equation}\label{acoustic.pot1sys.Jop}
\Jop = -\begin{pmatrix} 0 & D_t^{-1} \\ D_t^{-1} & 0 \end{pmatrix} .
\end{equation}

The equations in the potential system have the following variational structure
in terms of the inverse of the Noether operator \eqref{acoustic.pot1sys.Jop}, 
\begin{equation}\label{acoustic.pot1sys.Hop}
\Jop^{-1} = - \begin{pmatrix} 0 & D_t \\ D_t & 0 \end{pmatrix}  .
\end{equation}
One sees that 
\begin{equation}
(p_r,u_r)^\t 
= -\Jop^{-1}(\partial_p E,\partial_u E)^\t
\end{equation}
where 
\begin{equation}
E = r^2 (\tfrac{1}{2}p^2 -\tfrac{1}{3}\beta p^3) + r^{-2}\tfrac{1}{2} u^2 .
\end{equation}
This structure is, formally, a Hamiltonian formulation
in which $\Jop^{-1}$ and $E$ respectively play the roles of 
the Hamiltonian operator and the Hamiltonian density, 
where $r$ represents the ``time'' coordinate for the evolution 
and $t$ represents the ``spatial'' coordinate with respect to which 
$\Jop^{-1}$ is skew-adjoint. 
Also, $E$ is in fact the density of the energy integral \eqref{acoustic.ener}
expressed in terms of $p$ and $u$: 
$\mathcal E = \int_0^\infty \big( r^2 (\tfrac{1}{2}p^2 -\tfrac{1}{3}\beta p^3) + r^{-2}\tfrac{1}{2} u^2 \big)\,dr$.

\section{Concluding remarks}\label{sec:remarks}

For general PDE systems, 
the general results developed in \Ref{Anc2022a} 
on the basic algebraic structure surrounding adjoint-symmetries and symmetry actions
are very rich. 
In particular, 
as shown by the examples of physically interesting PDE systems in the present work, 
the adjoint-symmetry bracket constitutes a homomorphism of 
a Lie (sub) algebra of symmetries into a Lie algebra of adjoint-symmetries,
which can hold, surprisingly, even for dissipative systems with no variational structure. 
Moreover, whenever a PDE system possesses a non-multiplier adjoint-symmetry, 
there a dual symmetry action that yields a Noether operator
which leads to existence of a variational structure (Hamiltonian or Lagrangian). 

Thus, the adjoint-symmetries of a given PDE system 
carry useful information about important aspects of the system. 
Further developments and exploration of more examples 
will be an interesting problem for future work.

\section*{Acknowledgments}
SCA is supported by an NSERC Discovery Grant. 

This paper is dedicated to George Bluman: 
a wonderful colleague, research collaborator, and book co-author, 
who led me on my first steps into the beautiful subject of symmetries and their applications to differential equations. 
The results bring a full circle to the very fruitful work that we started together on multipliers and adjoint-symmetries more than two decades ago \cite{AncBlu1997a}.

\end{document}